# Lindstedt series, ultraviolet divergences and Moser's theorem

F. Bonetto$^\bullet$, G. Gallavotti$^*$, G. Gentile$^\dagger$, V. Mastropietro$^\star$

$^\bullet$ Matematica, Università di Roma, P.le Moro 2, 00185 Roma
$^*$ Fisica, Università di Roma, P.le Moro 2, 00185 Roma
$^\dagger$ IHES, 35 Route de Chartres, 91440 Bures sur Yvette
$^\star$ Matematica, Università di Roma II, Via Ricerca Scientifica, 00133 Roma

ABSTRACT.

*Moser's invariant tori for a class of nonanalytic quasi integrable even hamiltonian systems are shown to be analytic in the perturbation parameter. We do so by exhibiting a summation rule for the divergent series ("Lindstedt series") that formally define them. We find additional cancellations taking place in the formal series, besides the ones already known and necessary in the analytic case (i.e. to prove convergence of Lindtsedt algorithm for Kolmogorov's invariant tori). The method is interpreted in terms of a non renormalizable quantum field theory, considerably more singular than the one we pointed out in the analytic case.*

## 1. Introduction

**1.1. Hamiltonian. Tori and functional equations.** We consider a model ("Thirring model") with hamiltonian:

$$\mathcal{H} = \frac{1}{2J}\mathbf{A}\cdot\mathbf{A} + \varepsilon f(\boldsymbol{\alpha}), \qquad f(\boldsymbol{\alpha}) = \sum_{\boldsymbol{\nu}\in\mathbb{Z}^\ell} e^{i\boldsymbol{\nu}\cdot\boldsymbol{\alpha}} f_{\boldsymbol{\nu}}, \qquad f_{\boldsymbol{\nu}} = f_{-\boldsymbol{\nu}}, \qquad (1.1)$$

which can be interpreted as a model for $\ell$ interacting rotators if $\mathbf{A}\in\mathbb{R}^\ell, \boldsymbol{\alpha}\in\mathbb{T}^\ell$ are the angular momenta and angular positions of the rotators and $J$ is a positive "inertia moments" matrix.

Let $\boldsymbol{\omega}_0 \equiv J^{-1}\mathbf{A}_0$ be a rotation vector satisfying a *diophantine condition*. Kolmogorov's theorem, [K], states, for $\varepsilon$ small, the existence of a one parameter family $\varepsilon \to \mathcal{T}_\varepsilon$ of tori with parametric equations:

$$\mathbf{A}(\boldsymbol{\psi}) = \mathbf{A}_0 + \mathbf{H}(\boldsymbol{\psi}), \qquad \boldsymbol{\alpha}(\boldsymbol{\psi}) = \boldsymbol{\psi} + \mathbf{h}(\boldsymbol{\psi}), \qquad \boldsymbol{\psi}\in\mathbb{T}^\ell. \qquad (1.2)$$

If $f(\boldsymbol{\alpha})$ is an analytic function it is known that $\mathbf{H}$ and $\mathbf{h}$ are analytic functions of $\varepsilon$; this consequence of Kolmogorov's theorem was proved in this form by Moser, [M3], and, recently, by Eliasson, [E], by showing the convergence of the Lindstedt series. Our aim is to study the analitycity in $\varepsilon$ of $\mathbf{H}$ and $\mathbf{h}$ if $f \in C^{(p)}(\mathbb{T}^\ell)$ for some $p$ large enough. We have from other Moser's works, [M1,M2], (see also [M4,M5]), the existence of $\mathbf{H}$ and $\mathbf{h}$, (Moser's theorem), but no results seem to be known about the analyticity in $\varepsilon$: this is probably because the classical Moser's works rely on implicit functions theorems and other tools, which are essentially real analysis tools.

The problem is equivalent, see [G2], to studying the following functional equation:

$$\mathbf{h}(\boldsymbol{\psi}) = -\frac{1}{J}\varepsilon(\boldsymbol{\omega}_0 \cdot \partial_{\boldsymbol{\psi}})^{-2} \partial_\alpha f(\boldsymbol{\psi} + \mathbf{h}(\boldsymbol{\psi})) \,, \tag{1.3}$$

where $\partial_\alpha$ denotes derivative with respect to the argument.

**1.2. A simpler equation.** We start by considering the easier problem of solving the functional equation:

$$\mathbf{h}(\boldsymbol{\psi}) = \varepsilon \partial_\alpha f(\boldsymbol{\psi} + \mathbf{h}(\boldsymbol{\psi})) \,, \qquad f \text{ even} \,, \tag{1.4}$$

which is equivalent to "ignore" the small divisors problem. Moreover the existence of an analytic solution of (1.4) is a (somewhat non trivial, at least if $\ell > 1$, from our expansions viewpoint and) interesting problem by itself.

**1.3. Perturbation classes.** Our results concern functions $f$ in the class $\hat{C}^{(p)}(\mathbb{T}^\ell)$ which we define to be the functions $f(\boldsymbol{\alpha})$ that can be written as in (1.1) with $f_{\mathbf{0}} = 0$ and for $\boldsymbol{\nu} \neq \mathbf{0}$, $|\boldsymbol{\nu}| = \sqrt{\boldsymbol{\nu} \cdot \boldsymbol{\nu}}$, $||\boldsymbol{\nu}|| = \sum_{j=1}^\ell |\nu_j|$:

$$f_{\boldsymbol{\nu}} = \sum_{n \geq p+\ell}^{N} \frac{c_n + d_n(-1)^{||\boldsymbol{\nu}||}}{|\boldsymbol{\nu}|^n} \,, \tag{1.5}$$

for some $N > 0$. Note that all such functions have an *even Fourier transform*: a property that will be heavily used. See the remarks at the end of §2, and in §7. The quantity $|\boldsymbol{\nu}|$ will be always distinguished from $||\boldsymbol{\nu}||$.

For the purpose of comparison with more general cases we denote $\hat{C}^{(p)}_{\text{odd}}(\mathbb{T}^\ell)$ the functions that can be written as in (1.5) with $|\boldsymbol{\nu}|^{-n}$ replaced by $(\mathbf{u}_n \cdot \boldsymbol{\nu})|\boldsymbol{\nu}|^{-n-1}$ with $\mathbf{u}_n$ a unit vector.

The main results can be easily extended to functions somewhat more general than the above, see concluding remarks, and in their simplest form are summarized in the following two theorems, (here $C^{(<q)}$ means $C^{(q')}$ for all $q' < q$).

**1.4. Theorem.** *The equation (1.4) with $f \in \hat{C}^{(p)}(\mathbb{T}^\ell)$ admits a $C^{(<p-2)}$ solution analytic in $\varepsilon$ for $|\varepsilon|$ small enough if $p > 2$.*

Note that for $\ell = 1$ (1.4) with $f \in \hat{C}^{(p)}(\mathbb{T})$ and (1.3) with $f \in \hat{C}^{(p-2)}(\mathbb{T})$ are very closely related (although, strictly speaking, *not* equivalent because of the nonlinearity).

**1.5. Theorem.** *Suppose that $\boldsymbol{\omega}_0$ verifies a diophantine condition: $|\boldsymbol{\omega}_0 \cdot \boldsymbol{\nu}| \geq C_0^{-1}|\boldsymbol{\nu}|^{-\tau}$ for some $C_0, \tau > 0$ and for $\boldsymbol{\nu} \in \mathbb{Z}^\ell, \boldsymbol{\nu} \neq \mathbf{0}$. Then the equation (1.3) with $f \in \hat{C}^{(p)}(\mathbb{T}^\ell)$ admits a $C^{(0)}(\mathbb{T}^\ell)$ solution analytic in $\varepsilon$ for $|\varepsilon|$ small enough if $p > 3 + 6\tau$. In fact the regularity of the solution is $C^{(<p-3-6\tau)}$ for $p > 3 + 6\tau$.*

In both theorems the functions $\mathbf{h}$ are odd in $\boldsymbol{\psi}$ and divisible by $\varepsilon$. Theorem 1.4 is likely to be optimal (as discussed at the end of §4 and in spite of the weaker implication it yields when applied to the example in §2), while Theorem 1.5 is considerably weaker than the best known results on Moser's theorem, see [H,M2], in comparable classes of regularity for $f$; but we have not attempted at finding the best results that our technique permits, see comments in §7.

In this paper we shall assume that $\boldsymbol{\omega}_0$ (fixed once and for all) verifies the diophantine condition in Theorem 1.5: but to avoid carrying around too many constants we define $\boldsymbol{\omega} = C_0 \boldsymbol{\omega}_0$ and redefine $f$ to denote $C_0^2 f$ so that the dipohantine condition can be written:

$$|\boldsymbol{\omega} \cdot \boldsymbol{\nu}| > |\boldsymbol{\nu}|^{-\tau}, \qquad \text{for all } \boldsymbol{\nu} \in \mathbb{Z}^\ell, \boldsymbol{\nu} \neq \mathbf{0}, \tag{1.6}$$

and (1.3) can be written in the same form with $\boldsymbol{\omega}$ replacing $\boldsymbol{\omega}_0$ (with the new meaning of $f$).

## 2. Examples

The functional equation (1.4) admits for $\varepsilon$ small a $C^{(0)}$ solution if $f \in C^{(2)}$, even or not even, because the map $\mathbf{h} \to \mathbf{h}' = \varepsilon \partial_\alpha f(\boldsymbol{\psi} + \mathbf{h}(\boldsymbol{\psi}))$ is a "contraction" in $C^{(0)}$: $|\varepsilon \partial_\alpha f(\boldsymbol{\psi} + \mathbf{h}_1) - \varepsilon \partial_\alpha f(\boldsymbol{\psi} + \mathbf{h}_2)| \leq |\varepsilon| \, ||f||_{C^{(2)}} ||\mathbf{h}_1 - \mathbf{h}_2||_{C^{(0)}}$. In fact if $f \in C^{(p)}$ then the map is a contraction in $C^{(p-2)}$ for $||\varepsilon f||_{C^{(p)}} < 1$. Therefore if $f \in C^{(p)}$, $p \geq 2$, then $\mathbf{h}$ exists for $\varepsilon$ small and it is of class $C^{(p-2)}$. Here one can replace $C^{(p)}$ with $C^{(p-1)}$ *and* $(p-1)$th derivatives verifying a Lipshitz condition, a space that can be denoted $\text{Lip}_p(\mathbb{T}^\ell)$. If $f$ is even then the solution $\mathbf{h}$ is odd because in this case the space of the odd functions is invariant under the above map.

In this section we give two examples in which $\ell = 1$ and the functional equation:

$$h(\psi) = \varepsilon \partial f(\psi + h(\psi)), \qquad \psi \in \mathbb{T}^1, \tag{2.1}$$

($\partial = \partial_\alpha$) can be explicitly solved.

**2.1. Odd perturbations.** The simplest case corresponds to an odd $f$:

$$f(\psi) = \frac{1}{2} \psi(|\psi| - \pi), \qquad |\psi| \leq \pi, \tag{2.2}$$

as a periodic function of period $2\pi$ and zero average. This function belongs to $C^{(<2)}(\mathbb{T}^1)$ (and to $\text{Lip}_2(\mathbb{T}^1)$) and, with the notations in §1.3, to $\hat{C}^{(2)}_{\text{odd}}(\mathbb{T}^1)$ because its Fourier trasform $f_\nu$ is proportional to $i[(-1)^\nu - 1]\nu^{-3}$, $\nu \neq 0$, and $f_0 = 0$. The first step of an iterative method to solve (2.1) gives, as approximate solution, $\varepsilon \partial f(\psi + \varepsilon \partial f(\psi)) = \varepsilon(|\psi + \varepsilon(|\psi| - \frac{\pi}{2})| - \frac{\pi}{2})$, which for $\psi = 0$ reduces to $\varepsilon(|\varepsilon| - 1)\pi/2$, which is therefore non analytic in $\varepsilon$. Moreover it is easy to see that the exact solution of (2.1) can be written as:

$$\begin{aligned} h(\psi) &= h_+(\psi)\,\chi(\psi \in [\varepsilon\tfrac{\pi}{2}, (1 - \tfrac{\varepsilon}{2})\pi]) + h_-(\psi)\,\chi(\psi \in [-(1 + \tfrac{\varepsilon}{2})\pi, \varepsilon\tfrac{\pi}{2}]), \\ h_+(\psi) &= \frac{\varepsilon}{1 - \varepsilon}\left(\psi - \frac{\pi}{2}\right), \qquad h_-(\psi) = -\frac{\varepsilon}{1 + \varepsilon}\left(\psi + \frac{\pi}{2}\right), \end{aligned} \tag{2.3}$$

where $\chi(P)$ is the characteristic function of the set of points satisfying the condition $P$. So $h \in C^{(<1)}(\mathbb{T}^1)$, (hence $h \in C^{(0)}$ if one considers only integer derivatives), but it is not analytic in $\varepsilon$ for $\varepsilon$ small.

**2.2. Even perturbations.** Consider now the function (with derivative proportional to (2.2)):

$$f(\psi) = \frac{|\psi|^3}{3} - \frac{\psi^2 \pi}{2} + \frac{\pi^3}{12}, \qquad |\psi| \leq \pi, \tag{2.4}$$

viewed as a periodic function of period $2\pi$ and zero average; this function belongs to $C^{(<3)}(\mathbb{T}^1)$ (and to $\text{Lip}_3(\mathbb{T}^1)$) and its Fourier trasform $f_\nu$ is 0 for $\nu = 0$ and it is proportional to $[(-1)^\nu - 1]\nu^{-4}$ for $\nu \neq 0$; so that $f \in \hat{C}^{(3)}(\mathbb{T}^1)$. Note that:

$$\partial f(\psi + \varepsilon \partial f(\psi)) = (\psi + \varepsilon \psi(|\psi| - \pi))\left[|\psi + \varepsilon \psi(|\psi| - \pi)| - \pi\right], \tag{2.5}$$

which is analytic in $\varepsilon$ for $|\varepsilon| < 1$ because $\text{sign}(\psi) = \text{sign}(\psi + \varepsilon \psi(|\psi| - \pi))$ for every $|\psi| \leq \pi$ and $|\varepsilon| < 1$.

To solve the full equation $h(\psi) = \varepsilon(\psi + h(\psi))[|\psi + h(\psi)| - \pi]$, we consider the two possible cases:
(1) $\psi + h > 0$: there are two solutions of the second order equation for $h$, but only one which is $O(\varepsilon)$ for $\varepsilon \to 0$; so we fix:

$$h_+ = (2\varepsilon)^{-1}\left\{1 + \varepsilon\pi - 2\varepsilon\psi - \left((1+\varepsilon\pi)^2 - 4\varepsilon\psi\right)^{1/2}\right\} \tag{2.6}$$

Note that $\psi + h_+(\psi) > 0$ for $\psi > 0$.
(2) $\psi + h < 0$: like in the previous case there are two solutions but only the one with the plus sign is $O(\varepsilon)$ for $\varepsilon \to 0$; so we fix:

$$h_- = -(2\varepsilon)^{-1}\left\{1 + \varepsilon\pi + 2\varepsilon\psi - \left((1+\varepsilon\pi)^2 + 4\varepsilon\psi\right)^{1/2}\right\} \tag{2.7}$$

One has $\psi + h_-(\psi) > 0$ for $\psi < 0$.
It follows that the solution which is uniformly of order $O(\varepsilon)$ can be written as:

$$h(\psi) = h_+(\psi)\,\chi(\psi \in [0, \pi]) + h_-(\psi)\,\chi(\psi \in [-\pi, 0])\,, \tag{2.8}$$

which is analytic in $\varepsilon$ for $\varepsilon$ small enough, because $h_-(0) = h_+(0)$, and both $h_-$ and $h_+$ are analytic. One checks that $h \in C^{(<2)}(\mathbb{T}^1) \cap \hat{C}^{(2)}(\mathbb{T}^1)$ ($< 2$ becomes 1, i.e. $f \in C^{(1)}$, if one insists on integer order derivatives).

**2.3. Remarks.** One checks that, in general, if $f$ is odd and $\partial_\alpha f(0) \neq 0$ then the equation (2.1) does not admit a solution which is analytic in $\varepsilon$ for small $\varepsilon$. Nevertheless we have seen (by the contraction principle at the beginning of this section applied to the space $C^{(p)}$, forgetting the parity properties) that if $f \in C^{(p)}$ then (2.1) admits a $C^{(p-2)}$ solution for $\varepsilon$ small enough.

Therefore we see that the analyticity in $\varepsilon$ is linked in a non trivial way to the regularity of $f$. The functions in $\hat{C}^{(p)}$, i.e. verifying (1.5), have very special properties (being the kernels of homogeneous pseudodifferential operators on $\mathbb{T}^\ell$): namely they are regular (real analytic) for $\psi$ such that $\partial_j f(\psi) \neq 0$, (see, for instance, [SW] p. 256, which yields $C^{(\infty)}$ regularity for $\psi \neq \mathbf{0}$ and $|\psi_j| < \pi$; while analyticity can be seen, for instance, as a consequence of Theorem 1.4, see §7.4 below).

More generally we can conjecture that analyticity arises if $\partial f$ vanishes at the singularities of $f$, when $f$ is differentiable enough. In this sense the parity condition, together with (1.5), is a simple and rather general way of characterizing functions with the latter property. It is not difficult to set up assumptions under which this statement can be made precise and proved in an elementary way at least for $\ell = 1$. Such an approach could not be easily extended to the general case of (1.3) or even to the case of (1.4) with $\ell > 1$: hence in §3, §4 we prove Theorem 1.4 with a technique that admits a "straightforward extension" to the proof of Theorem 1.5.

As we shall see in the coming sections the property of $f$ of being even and verifying (1.5) (hence such that $\partial f$ vanishes at the singularities of $f$) is the root of the new cancellation mechanism that will permit us to deduce our results.

## 3. Formal solutions: LNP series.

We study now the equations (1.3) or (1.4) with $f_{\boldsymbol{\nu}} = |\boldsymbol{\nu}|^{-b}$ (with $b$ large enough) so that $f \in \hat{C}^{(b-\ell)}(\mathbb{T}^\ell)$. It will appear from the following proof that everything works quite more generally, and in particular for all $f \in \hat{C}^{(p)}$ (with $p$ large enough). We prove that $\mathbf{h}$ is analytic in $\varepsilon$, for $\varepsilon$ small, if $b$ is large enough.

By an argument analogous to [E,G2,GM1], we see that one can solve formally the equation, by writing the Fourier coefficients $\mathbf{h}_{\boldsymbol{\nu}}$, $\boldsymbol{\nu} \in \mathbb{Z}^\ell$, of $\mathbf{h}(\boldsymbol{\psi})$ as a power series in $\varepsilon$:

$$\mathbf{h}(\boldsymbol{\psi}) = \sum_{\boldsymbol{\nu} \in \mathbb{Z}^\ell} \mathbf{h}_{\boldsymbol{\nu}} e^{i\boldsymbol{\nu}\cdot\boldsymbol{\psi}}, \qquad \mathbf{h}_{\boldsymbol{\nu}} = \varepsilon \mathbf{h}_{\boldsymbol{\nu}}^{(1)} + \varepsilon^2 \mathbf{h}_{\boldsymbol{\nu}}^{(2)} + \dots, \qquad (3.1)$$

(*Lindstedt-Newcomb-Poincaré series*, or *LNP series*, or *Lindstedt series*).

The rules to construct the expansion for the case (1.4) are identical to those discussed in [G2] for the equation (1.3), except that the small divisors $(i\boldsymbol{\omega}\cdot\boldsymbol{\nu})^{-2}$ are replaced by 1, and $J$ by $-1$.

The expansion is purely formal: this is so in [E,G2] only because of the "small divisors" problem. Here it is formal also because it involves computing arbitrarily high derivatives of $f$ (while $f$ is only supposed to possess few derivatives).

For completeness we recall, briefly, the algorithm rules for the expansion: the generalization to (1.4) is a trivial adaptation of the algorithm leading to the Lindstedt series, amply described in the literature, see for instance Appendix R in [G2] and [GM3]. Hence we just describe it without comments.

**3.1. The graph representation (polynomial case).** We begin by supposing that $f$ is a trigonometric polynomial. It is necessary to recall, first, the notion of rooted graph as used here. We lay down one after the other, on a plane, $k$ pairwise distinct unit segments oriented from one extreme to the other: respectively the *initial point* and the *endpoint* of the oriented segment. The oriented segment will also be called *arrow*, *branch* or *line*. The segments are supposed to be numbered from 1 to $k$.

The rule is that after laying down the first segment, the *root branch*, with the endpoint at the origin and otherwise arbitrarily, the others are laid down one after the other by attaching an endpoint of a new branch to an initial point of an old one and by leaving free the new branch initial point. The set of initial points of the object thus constructed will be called the set of the graph *nodes* or *vertices*. A graph of *order k* is therefore a partially ordered set of $k$ nodes with top point the endpoint of the root branch, also called the *root* (which is not a node); there will be several "bottom nodes", unless the graph is a succession of lines each attached to the previous one (the latter case, *linear graph*, although trivial will be a very important one). Therefore the graphs are "trees" in the sense of graph theory. If we take into account also the numbers labeling the branches, we have "numbered graphs".

It is perhaps worth commenting explicitly that with the above conventions the root of the graph is at the top while the bottom points are at the bottom *in the sense of the partial order on the graph generated by the arrows directions*: this may be sometimes confusing as it is contrary to the intuition if one thinks of drawing the graph as a real life tree looks like. On the other hand *this is a natural convention* if one thinks that the arrows that constitute the branches, and define the partial order on the graph, point towards the root. We denote by $\leq$ the ordering relation, and say that two nodes $v$, $w$ are "comparable" if $v < w$ or $w < v$.

With each graph node $v$ we associate an *external momentum* or *mode* which is simply an integer component vector $\boldsymbol{\nu}_v \neq \mathbf{0}$; with the root of the graph (which is not regarded as a node) we associate a label $j = 1, \ldots, \ell$. The labels attached to the graphs (the numbers being included) will be referred as *decorations*.

For each node $v$, we denote by $v'$ the node immediately following $v$ and by $\lambda_v \equiv v'v$ the branch connecting $v$ to $v'$, ($v$ will be the initial point and $v'$ the endpoint of $\lambda_v$). If $v$ is the node immediately preceding the root $r$ (*highest node*) then we shall write $v' = r$, for uniformity of notation (recall that $r$ is not a node). We consider "comparable" two lines $\lambda_v$, $\lambda_w$, if $v$, $w$ are such.

Given a graph $\vartheta$ let $p_v$ be the number of branches entering the node $v$: then each of the $p_v$ branches can be thought as the root branch of a *subgraph* having root at $v$: the subgraph is uniquely determined by $v$ and one of the $p_v$ nodes $w$ immediately preceding $v$. Hence if $w' = v$ it can be denoted $\vartheta_{vw}$. It is useful also to consider graphs $\vartheta^0$ bearing no momentum labels and we use the notation $\vartheta = (\vartheta^0, \{\boldsymbol{\nu}_x\})$; the subtrees with no momentum labels are denoted by $\vartheta^0_{vw}$.

The angles at which the segments are attached will be irrelevant, *i.e.* the operation of changing the angles between arrows emerging from the same node (each arrow carrying along, unchanged, the subgraph of arrows possibly attached to its initial point) generates a group of transformations, and two graphs that can be overlapped by acting on them with a group element are regarded as identical.

We can also introduce another group of transformations, which consist of permuting the subgraphs entering into a node $v$, and we consider equivalent graphs which can be overlapped by acting on them with a group element in such a way that all the labels match. The number of (non equivalent numbered) graphs with $k$ branches is thus bounded by $4^k k!$, [HP].

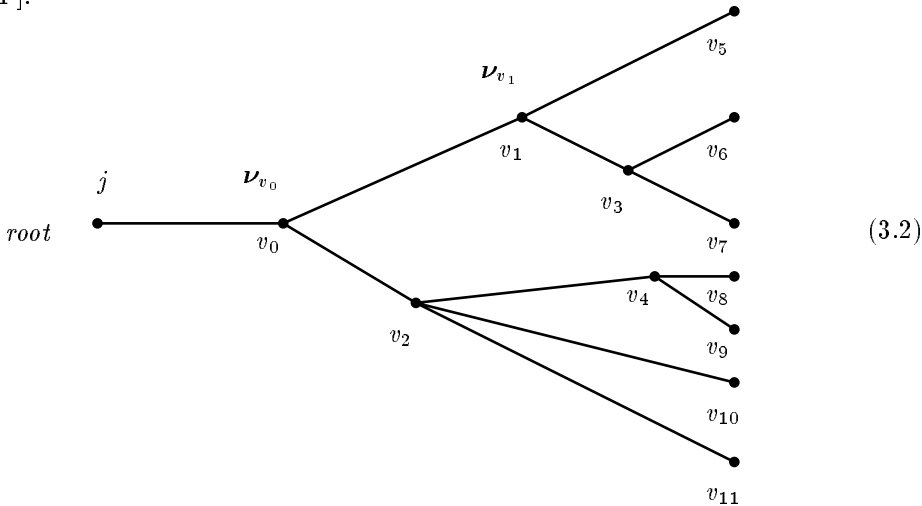

(3.2)

**Fig.3.1**. A graph $\vartheta$ with $p_{v_0} = 2, p_{v_1} = 2, p_{v_2} = 3, p_{v_3} = 2, p_{v_4} = 2$ and $k = 12$, $\prod p_v! = 2^4 \cdot 6$, and some decorations. The line numbers, distinguishing the lines, and the arrows, pointing at the root, are not shown. The lines length should be 1 but it is drawn of arbitrary size.

For each node $v$, we denote by $v'$ the node immediately following $v$ and by $\lambda_v = v'v$ the branch connecting $v$ to $v'$, ($v$ will be the initial point and $v'$ the endpoint of $\lambda_v$). If $v$ is the node immediately preceding the root $r$ (*highest node*) then we shall write $v' = r$, for uniformity of notation (recall that $r$ is not a node). We consider "comparable" two lines $\lambda_v$, $\lambda_w$, if $v$, $w$ are such.

Given a graph $\vartheta$ let $p_v$ be the number of branches entering the node $v$: then each of the $p_v$ branches can be thought as the root branch of a *subgraph* having root at $v$: the subgraph is uniquely determined by $v$ and one of the $p_v$ nodes $w$ immediately preceding $v$. Hence if $w' = v$ it can be denoted $\vartheta_{vw}$. It is useful also to consider graphs $\vartheta^0$ bearing no momentum labels and we use the notation $\vartheta = (\vartheta^0, \{\nu_x\})$; the subtrees with no momentum labels are denoted by $\vartheta^0_{vw}$.

With each branch $\lambda_v$, we associate another integer component vector, the *branch momentum* "flowing through the branch", defined by $\nu_{\lambda_v} = \sum_{w \leq v} \nu_w$. Then, given a positive matrix $J$ and a trigonometric polynomial $f(\psi) = \sum_{0 < |\nu| \leq N} f_\nu \cos \nu \cdot \psi$, $f_\nu = f_{-\nu}$, we consider from now on only graphs $\vartheta$ with $k$ branches, "decorated" by node labels $\nu_v$ *such that $\nu_\lambda \neq 0$ for all $\lambda$*, and associate with each decorated graph the *value*

$$\text{Val}(\vartheta) = -i \prod_{v < r} f_{\nu_v} \frac{\nu_{v'} \cdot J^{-1} \nu_v}{(i\boldsymbol{\omega} \cdot \boldsymbol{\nu}_{\lambda_v})^X} \;, \tag{3.3}$$

where $v'$ is the node immediately following $v$ in $\vartheta$; here $\nu_r$ denotes the unit vector in the $j$th direction, $\nu_r = \mathbf{e}_j$, $j = 1, \ldots, \ell$, and $X = 0, -J = 1$ for (1.4) or $X = 2$ for (1.3). When $X = 2$, $[i\boldsymbol{\omega} \cdot \boldsymbol{\nu}_{\lambda_v}]^{-2}$ will be called the *divisor* of the line $\lambda_v$. To stress the (deep) analogy with quantum field theory (QFT, see §7), in which our graphs play the role of *Feynman graphs*, (see §7), we shall sometimes also call $g_\lambda \equiv [i\boldsymbol{\omega} \cdot \boldsymbol{\nu}_\lambda]^{-2}$ the *propagator* of the line $\lambda$.

The momentum flowing through the root will be denoted also $\nu(\vartheta)$. The *Lindstedt-Newcomb-Poincaré* ("LNP") polynomial $\mathbf{h}^{(k)}(\psi)$ is *defined* by $\sum_\nu \mathbf{h}^{(k)}_\nu e^{i\nu \cdot \psi}$, with:

$$\begin{aligned} h^{(k)}_{j\nu} &= \frac{1}{k!} \sum_{\vartheta, \, \nu(\vartheta) = \nu} \text{Val}(\vartheta) = \frac{1}{k!} \sum_{\vartheta^0} W(\vartheta^0, \nu) \\ W(\vartheta^0, \nu) &\stackrel{def}{=} \sum_{\nu(\vartheta) = \nu} \text{Val}(\vartheta). \end{aligned} \tag{3.4}$$

Hence, *in the case of a trigonometric polynomial $f$*, the only difference between the formal solutions to (1.3) and (1.4), besides the value of $J$, is that in the second case $X = 0$ so that *no (small) divisors appear* in the solution to (1.4).

Kolmogorov's theorem yields that, if $\boldsymbol{\omega}$ verifies the diophantine condition (1.6) for some $\tau > 0$ and $f$ is a trigonometric polynomial, then:

**3.2. Theorem.** *Given the hamiltonian system (1.1), where $f(\boldsymbol{\alpha})$ is a trigonometric polynomial in the angle variables, i.e. $\boldsymbol{\nu} \in \mathbb{Z}^\ell$ with $|\boldsymbol{\nu}| \leq N$ for some integer $N$, then the Lindstedt series defined through (3.1) is convergent for $\varepsilon$ small enough.*

The theorem still holds under weaker assumptions on the perturbation, (see §3.3), and it is a consequence of Kolmogorov's theorem, (see [M3]); it has been recently proved with new techniques by Eliasson, (see [E]). The corresponding statement for (1.4) is also correct (and trivially so).

**3.3. The differentiable case. Abel's summation.** The new techniques are, in fact *not* restricted neither to polynomial $f$ (analytic is sufficient) nor to even $f$. If $f$ is analytic in $\boldsymbol{\alpha}$ the above algorithms still make sense: the sums over the incoming momenta are simply no longer restricted to have a bounded size. This implies that the sum in (3.4) is now a genuine infinite sum, which however is trivially absolutely convergent in the case of (1.4). In fact the coefficients $f_{\boldsymbol{\nu}}$ decay exponentially as $|\boldsymbol{\nu}| \to \infty$.

In the case of (1.3) the convergence (still at $k$ fixed) is also trivial because of the supposed diophantine condition enjoyed by the rotation vector $\boldsymbol{\omega}$.

In both cases also the convergence of the sum over $k$ with weight $\varepsilon^k$ can be established (see for instance [E,CF,G2,GM2,GM3]).

But the situation is remarkably different in the case of differentiable $f$, e.g. when $f \in \hat{C}^{(p)}(\mathbb{T}^\ell)$: to proceed we need, therefore, a *summation rule* for the formal series that arise in the Lindstedt algorithm.

For each node $v \in \vartheta$, we can write $f_{\boldsymbol{\nu}_v} = f_{\boldsymbol{\nu}_v} e^{-\kappa |\boldsymbol{\nu}_v|}$, where $\kappa$ should be zero. *The summation rule is that the parameter $\kappa$ (ultraviolet cut off) should be taken $\kappa > 0$: after computing the coefficients $\mathbf{h}_{\boldsymbol{\nu},\kappa}^{(k)}$ one will perform the limit as $\kappa \to 0$ which will define $\mathbf{h}_{\boldsymbol{\nu}}^{(k)}$.* The summation rule could be called *Abel's summation* of the Lindstedt series coefficients.

The next section is devoted to the proof that, in the case of (1.4), the latter limit exists and defines the Fourier coefficients of a function $\mathbf{h}$ which has class, at least, $C^{(0)}$ and such that the series (3.1) converges in the norm of $C^{(0)}$ if $f \in \hat{C}^{(p)}$ with $p$ large. In fact we prove that

$$\sum_{\boldsymbol{\nu} \in \mathbb{Z}^\ell} |\boldsymbol{\nu}|^s |h_{j\boldsymbol{\nu}}^{(k)}| = \sum_{\boldsymbol{\nu} \in \mathbb{Z}^\ell} \frac{1}{k!} \Big| \sum_{\vartheta,\, \boldsymbol{\nu}(\vartheta)=\boldsymbol{\nu}} |\boldsymbol{\nu}(\vartheta)|^s \mathrm{Val}(\vartheta) \Big| = \sum_{\boldsymbol{\nu} \in \mathbb{Z}^\ell} \frac{1}{k!} \Big| \sum_{\vartheta^0} |\boldsymbol{\nu}|^s W(\vartheta^0, \boldsymbol{\nu}) \Big| \le C_0^k ,$$
(3.5)

for a suitable constant $C_0$, if $f \in C^{(p+s)}(\mathbb{T}^\ell)$, with $p \ge 2$, so that $\mathbf{h}$ turns out to be analytic in $\varepsilon$ for $|\varepsilon| < C_0^{-1} \equiv \varepsilon_0$.

## 4. Multiscale analysis for the model (1.4).
## Ultraviolet divergences.

For simplicity we suppose that $f_{\boldsymbol{\nu}} = |\boldsymbol{\nu}|^{-b}$: the extension to the more general (1.5) is straightforward.

**4.1. Divergences.** Given a graph $\vartheta$, we can define the *scale $h_v$* of the node $v$ to be the integer $h_v \ge 1$ such that $2^{h_v-1} \le |\boldsymbol{\nu}_v| < 2^{h_v}$. We say that the labels $\{\boldsymbol{\nu}_x\}$ and $\{h_x\}$ are *compatible* if $|\boldsymbol{\nu}_v| \in [2^{h_v-1}, 2^{h_v})$ for all $v \in \vartheta$. The compatibility relationship between $\{\boldsymbol{\nu}_x\}$ and $\{h_x\}$ will be denoted $\{\boldsymbol{\nu}_x\}$ comp $\{h_x\}$.

Then we can write in (3.5)

$$\sum_\vartheta \mathrm{Val}(\vartheta) = \sum_{\vartheta^0} \sum_{\{\boldsymbol{\nu}_x\}} \mathrm{Val}(\vartheta^0, \{\boldsymbol{\nu}_x\}) = \sum_{\vartheta^0} \sum_{\{h_x\}} \sum_{\{\boldsymbol{\nu}_x\} \,\mathrm{comp}\, \{h_x\}} \mathrm{Val}(\vartheta^0, \{\boldsymbol{\nu}_x\}) , \qquad (4.1)$$

and the resulting terms, in which every node has a definite "scale", are the addends of the *multiscale decomposition* of the original graph $\vartheta$. The labels $\{h_x\}$ will be called *ultraviolet scale labels* (to contrast them with the "infrared scale" labels of §5).

Given a graph $\vartheta^0$, a bound on the values $\mathrm{Val}(\vartheta^0, \{\boldsymbol{\nu}_x\})$ can be found immediately, from (3.3) and from the scaling properties of the $f_{\boldsymbol{\nu}_v}$, to be, if $\boldsymbol{\nu} = \boldsymbol{\nu}(\vartheta) \equiv \sum_v \boldsymbol{\nu}_v$:

$$\sum_{\{h_x\}} \sum_{\{\boldsymbol{\nu}_x\} \,\mathrm{comp}\, \{h_x\}} |\boldsymbol{\nu}|^s |\mathrm{Val}(\vartheta^0, \{\boldsymbol{\nu}_x\})| \le B^k \sum_{\{h_x\}} \prod_{v \in \vartheta} 2^{h_v(\ell+s+1+p_v-b)} , \qquad (4.2)$$

where $|\boldsymbol{\nu}| \leq \sum_v |\boldsymbol{\nu}_v| \leq 2^k \prod_v |\boldsymbol{\nu}_v|$ and $B$ is a suitable constant. It is important to note that the assignments of the labels $\{h_x\}$ in (4.2) allow us to say that $f_{\boldsymbol{\nu}_v}$ is bounded by $2^{-b(h_v-1)}$.

We call this kind of bounds *dimensional bounds* or *power counting bounds*, as customary in quantum field theory, see [G1,BG].

Setting $b = 2 + s + \ell + \mu$, *i.e.* taking $f \in \hat{C}^{(2+s+\mu)}$, with $\mu > 0$, and exploiting (as usual in such "field theory arguments", see [G1]) the (trivial) identity:

$$\sum_{v \leq v_1} h_v p_v = \sum_{v < v_1} h_{v'}, \qquad v_1 = \text{ highest node in } \vartheta^0, \tag{4.3}$$

(recall that we denote $v'$ the node immediately following $v$), one obtains from (4.2):

$$\sum_{\{h_x\}} \sum_{\{\boldsymbol{\nu}_x\} \text{ comp}\{h_x\}} |\boldsymbol{\nu}|^s |\text{Val}(\vartheta^0, \{\boldsymbol{\nu}_x\})| \leq \sum_{\{h_x\}} 2^{-h_{v_1}} \left( \prod_{v < v_1} 2^{h_{v'} - h_v} \right) \left( \prod_{v \in \vartheta} 2^{-\mu h_v} \right). \tag{4.4}$$

The sum over the ultraviolet labels $\{h_x\}$ in (4.4) can be bounded only if suitable relations are imposed on the scale labels, *e.g.* if $h_v \geq h_{v'}$, $\forall v \in \vartheta$; but this is not always the case, and we refer to the phenomenon as *ultraviolet divergences*: one has therefore to look for some cancellation mechanism between the values of different graphs which are source of the above divergences.

**4.2. The parity cancellations and interpolation.** Consider first the representation of $\mathbf{h}_{\boldsymbol{\nu}}^{(k)}$ by graphs $(\vartheta^0, \{\boldsymbol{\nu}_x\})$ without scale labels.

The cancellations will take place between values of graphs $(\vartheta^0, \{\boldsymbol{\nu}_x\})$ with the same $\vartheta^0$ and different external momenta assignments. In order to recognize which terms must be collected together to exhibit the cancellations, given a set of momenta and fixed a node $\bar{v} \in \vartheta^0$, we define the change of variables $U_{\bar{v}w}^{\sigma_w} : \mathbb{Z}^\ell \longleftrightarrow \mathbb{Z}^\ell$, where $w \in \mathcal{B}_{\bar{v}} = $ the set of the $p_{\bar{v}}$ nodes immediately preceding $\bar{v}$, by fixing a sign $\sigma_w = \pm 1$ and defining $U_{\bar{v}w}^{\sigma_w}(\{\boldsymbol{\nu}_x\}) = \{\boldsymbol{\nu}_x'\}$ as:

$$\begin{aligned}
\boldsymbol{\nu}_z' &= \sigma_w \boldsymbol{\nu}_z, \qquad z \geq w, \\
\boldsymbol{\nu}_z' &= \boldsymbol{\nu}_z, \qquad \text{for all other } z \neq \bar{v}, \\
\boldsymbol{\nu}_{\bar{v}}' &= \boldsymbol{\nu}_{\bar{v}} + (1 - \sigma_w) \sum_{z \leq w} \boldsymbol{\nu}_z \equiv \boldsymbol{\nu}_{\bar{v}} + (1 - \sigma_w) \boldsymbol{\nu}_{\lambda_w}.
\end{aligned} \tag{4.5}$$

The change of variables is one to one and, when it is not the identity (*i.e.* $\sigma_w = -1$), by acting on the external momentum labels it changes the graph $\vartheta = (\vartheta^0, \{\boldsymbol{\nu}_x\})$ to a graph $U_{\bar{v}w}^- \vartheta = (\vartheta^0, \{\boldsymbol{\nu}_x'\})$ such that all the momenta that flow through the corresponding lines of the two graphs are either the same (if the line follows $\lambda_{\bar{v}}$ or is not comparable to $\lambda_{\bar{v}}$) or change sign (otherwise). The momentum flowing through the root branch remains always the same. And all the scalar products of neighbouring nodes external momenta are the same in the two graphs *except* the product relative to the pairs of nodes $\bar{v}w$ or $\bar{v}'\bar{v}$ (where $\bar{v}' > \bar{v}$ is the node immediately following $\bar{v}$, see §3.1), which change from $\boldsymbol{\nu}_{\bar{v}} \cdot \boldsymbol{\nu}_w$ to $-(\boldsymbol{\nu}_{\bar{v}} + 2\boldsymbol{\nu}_{\lambda_w}) \cdot \boldsymbol{\nu}_w$ and, respectively, from $\boldsymbol{\nu}_{\bar{v}'} \cdot \boldsymbol{\nu}_{\bar{v}}$ to $\boldsymbol{\nu}_{\bar{v}'} \cdot (\boldsymbol{\nu}_{\bar{v}} + 2\boldsymbol{\nu}_{\lambda_w})$.

Note that the changes of variables $U_{v_1 w_1}^{\sigma_1}$ and $U_{v_2 w_2}^{\sigma_2}$ commute.

Therefore we can consider, given $(\vartheta^0, \{\boldsymbol{\nu}_x\})$ and fixed a node $w \in \mathcal{B}_{\bar{v}}$, $w' = \bar{v}$, the sum:

$$\sum_{\sigma = \pm 1} \text{Val}(\vartheta^0, U_{w'w}^\sigma \{\boldsymbol{\nu}_x\}), \tag{4.6}$$

and, more generally for any choice of the subset $\mathcal{B}_{1\bar{v}} \subseteq \mathcal{B}_{\bar{v}}$ of nodes immediately preceding $\bar{v}$, we can consider the sum:

$$\sum_{\substack{\sigma_w = \pm 1 \\ w \in \mathcal{B}_{1\bar{v}}}} \operatorname{Val}(\vartheta^0, \prod_{w \in \mathcal{B}_{1\bar{v}}} U_{\bar{v}w}^{\sigma_w}\{\boldsymbol{\nu}_x\}) \,. \tag{4.7}$$

The key remark is that the sum over the signs in (4.7) can be evaluated by an interpolation formula involving $|\mathcal{B}_{1\bar{v}}| \le p_{\bar{v}}$ auxiliary interpolation variables $t_w \in [0,1]$, $w \in \mathcal{B}_{1\bar{v}}$, because, as one can check after working out a few simple examples, the sum over the $\sigma_w$'s generates a composition of increments of $\operatorname{Val}(\vartheta^0, \{\boldsymbol{\nu}_x\})$. We express this as follows:

$$\begin{aligned}
\sum_{\{\sigma_w\}_{w \in \mathcal{B}_{1\bar{v}}}} \operatorname{Val}(\vartheta^0, \prod_{w \in \mathcal{B}_{1\bar{v}}} U_{\bar{v}w}^{\sigma_w}\{\boldsymbol{\nu}_x\}) &\equiv \\
\equiv \Big\{ \Big( \prod_{w \in \mathcal{B}_{1\bar{v}}} \int_1^0 dt_w \Big) \sum_{||\mathbf{a}_{\bar{v}}|| = p_{\bar{v}} + 1} & \Big( \prod_{w \in \mathcal{B}_{1\bar{v}}} \frac{\partial}{\partial t_w} \Big) \Big[ \big(i\boldsymbol{\nu}_{\bar{v}}(\mathbf{t}_{\bar{v}})\big)^{\mathbf{a}_{\bar{v}}} f_{\boldsymbol{\nu}_{\bar{v}}(\mathbf{t}_{\bar{v}})} \Big] \Big\} \cdot \operatorname{Val}' \,,
\end{aligned} \tag{4.8}$$

where:
(i) $\operatorname{Val}'$ is a tensor containing all the other value factors relative to nodes $v$'s different from $\bar{v}$.
(ii) The free indices of the tensor $\boldsymbol{\nu}_{\bar{v}}(\mathbf{t}_{\bar{v}})^{\mathbf{a}_{\bar{v}}}$ (which is of order $1 + p_{\bar{v}}$) are contracted (by performing the $\sum_{\mathbf{a}_{\bar{v}}}$) with the ones that appear in the tensor $\operatorname{Val}'$. And $\mathbf{a}_{\bar{v}}$ is a $\ell$ dimensional positive integer components vector (with $||\mathbf{a}_{\bar{v}}||$ denoting the sum of the components) and, given a vector $\mathbf{b}$, we put $\mathbf{b}^{\mathbf{a}} = b_1^{a_1} \ldots b_\ell^{a_\ell}$; furthermore $\mathbf{t}_{\bar{v}} = (t_{w_1}, \ldots, t_{w_{|\mathcal{B}_{1\bar{v}}|}})$ and $\boldsymbol{\nu}_{\bar{v}}(t_{w_1}, \ldots, t_{w_{|\mathcal{B}_{1\bar{v}}|}}) \equiv \boldsymbol{\nu}_{\bar{v}}(\mathbf{t}_{\bar{v}})$ is defined as:

$$\boldsymbol{\nu}_{\bar{v}}(\mathbf{t}_{\bar{v}}) = \boldsymbol{\nu}_{\bar{v}}(t_{w_1}, \ldots, t_{w_{|\mathcal{B}_{1\bar{v}}|}}) = \boldsymbol{\nu}_{\bar{v}} + \sum_{w \in \mathcal{B}_{1\bar{v}}} 2 t_w \, \boldsymbol{\nu}_{\lambda_w} = \boldsymbol{\nu}_{\bar{v}} + \sum_{w \in \mathcal{B}_{1\bar{v}}} t_w \Big( \sum_{z \le w} 2 \boldsymbol{\nu}_z \Big) \,, \tag{4.9}$$

where $\boldsymbol{\nu}_{\bar{v}}(\mathbf{t}_{\bar{v}}) = \boldsymbol{\nu}_{\bar{v}}$ if $\mathcal{B}_{1\bar{v}} = \emptyset$.

*The cancellations are expressed by the fact that the are no "initial" terms (corresponding to $t_w = 1$ in the integrals) in the interpolation formula (4.8) because the $\sum_U$ leads to a sum of increments: we call "new" such cancellations, in order to distinguish between them and the "old" cancellations which have to be exploited in the proof of the KAM theorem in the analytic case; the "old" cancellations will be discussed again in §5 below.*

In the value of a graph the external momentum $\boldsymbol{\nu}_{\bar{v}}$ of a node $\bar{v}$ appears in $f_{\boldsymbol{\nu}_{\bar{v}}}$ and as a factor in which it is raised to its $(p_{\bar{v}} + 1)$th power: each of the $p_{\bar{v}}$ entering lines contributes one power and one more power comes from the exiting line.

The external momenta correspond to derivatives of the perturbation $f$: hence one would say that in a graph in which there is a node with $p_v$ entering lines one needs, to make sense of it in the limit in which the ultraviolet cut off is removed, at least $(p_v + 1)$th differentiability of $f$.

REMARK. If one could *shift* each of the $p_v$ external momenta factors corresponding to the $p_v$ incoming lines to the *preceding nodes* $w_1, \ldots, w_{p_v}$ (with $w'_j = v$), then we would have in each node $v$ *at most 2 external momentum factors* (or 1 only, in the case of the highest node): *i.e.* if one could replace $|\boldsymbol{\nu}_v|^{p_v}$ by $\prod_{j=1}^{p_v} |\boldsymbol{\nu}_{w_j}|$ then we would have in each node just 2, or 1, external momentum factor. Hence we could hope to get away with $f$'s with just 2 derivatives.

One can check on the simplest graphs (*e.g.* on the graph with just two nodes) that the effect of the above interpolation is to *redistribute* the derivatives, that act on the $f$'s, among the various $f$'s corresponding to the nodes; thus producing precisely the net effect of shifting down (along the graph lines) the external momentum factors, so that formally one does not need more than 2 derivatives to make sense of the Lindstedt series coefficients when the ultraviolet cut off tends to 0. We recommend this calculation as it is very enlightening and it was in fact the key to our analysis.

(iii) The assumed form (1.5) of the $f_{\boldsymbol{\nu}}$ allows us to think that $f_{\boldsymbol{\nu}}$ is defined on $\mathbb{R}^{\ell}$ rather than on $\mathbb{Z}^{\ell}$ and hence to give a meaning to the derivatives of $f_{\boldsymbol{\nu}_v(\mathbf{t}_v)}$.

(iv) The integration over the $\mathbf{t}_{\bar{v}}$-variables may cause $\boldsymbol{\nu}_{\bar{v}}(\mathbf{t}_{\bar{v}})$ to pass through $\mathbf{0}$ where $f_{\boldsymbol{\nu}}$ is singular. In this case a convergence problem arises and we *must* treat it before being able to really use the interpolation formula (4.8): see comments after (4.21) below.

**4.3. Nodes out of order and interpolation.** From (3.4) we see that we should study the sum $S_k(\vartheta^0) = \sum_{\boldsymbol{\nu}} |\boldsymbol{\nu}|^s |W(\vartheta^0, \boldsymbol{\nu})|$ and by remembering that $\vartheta^0_{v_1 w}$ denote the subgraphs with root $v_1 = w'$ and highest node $w$ (see paragraph preceding (3.3)), we can write $S_k(\vartheta^0)$ as:

$$\begin{aligned} S_k(\vartheta^0) &= \sum_{\boldsymbol{\nu}} |\boldsymbol{\nu}|^s \Big| W(\vartheta^0, \boldsymbol{\nu}) \Big| \\ &= \sum_{\boldsymbol{\nu}} |\boldsymbol{\nu}|^s \Big| \prod_{w \in \mathcal{B}_{v_1}} \Big( \sum_{\boldsymbol{\nu}_{\lambda_w}} (i \boldsymbol{\nu}_{v_1})^{\mathbf{a}_{v_1}} f_{\boldsymbol{\nu}_{v_1}} W(\vartheta^0_{v_1 w}, \boldsymbol{\nu}_{\lambda_w}) \Big) \Big|, \end{aligned} \quad (4.10)$$

where $v_1$ is the highest node, $\boldsymbol{\nu}_{v_1} = \boldsymbol{\nu} - \sum_{w \in \mathcal{B}_{v_1}} \boldsymbol{\nu}_{\lambda_w}$ and $\mathcal{B}_v$ is the set of $p_v$ nodes that immediately precede $v$.

Fixed $\boldsymbol{\nu}$ and $\{\boldsymbol{\nu}_{\lambda_w}\}_{w \in \mathcal{B}_{v_1}}$ let $h_{v_1} = h_{v_1}(\boldsymbol{\nu}, \{\boldsymbol{\nu}_{\lambda_w}\}_{w \in \mathcal{B}_{v_1}})$ be *the scale of* $\boldsymbol{\nu}_{v_1}$: *i.e.* $\boldsymbol{\nu}_{v_1}$ is such that $2^{h_{v_1}-1} \leq |\boldsymbol{\nu}_{v_1}| < 2^{h_{v_1}}$. Given $w$ with $w' = v_1$ we say that $w$ is *out of order* with respect to $v$ if

$$2^{h_{v_1}} > 2^o p_{v_1} |\boldsymbol{\nu}_{\lambda_w}|, \qquad o = 5, \quad (4.11)$$

where $p_v$ is the number of branches entering $v$. It will become clear that the number 5 in (4.11) can be replaced with any integer $o \geq 5$: the actual value chosen for $o$ only influences the size or the ease of the bounds. We denote $\mathcal{B}_{1v_1} \equiv \mathcal{B}_{1v_1}(\boldsymbol{\nu}, \{\boldsymbol{\nu}_{\lambda_w}\}_{w \in \mathcal{B}_{v_1}}) \subseteq \mathcal{B}_{v_1}$ the nodes $w \in \mathcal{B}_{v_1}$ which are out of order with respect to $v_1$. The number of elements in $\mathcal{B}_{1v_1}$ will be denoted $q_v = |\mathcal{B}_{1v_1}|$. The notion of $w$ being out of order with respect to $v_1$ depends on $\{\boldsymbol{\nu}_{\lambda_w}\}_{w \in B_{v_1}}$ and $\boldsymbol{\nu}$.

Given a set $\{\boldsymbol{\nu}_{\lambda_w}\}_{w \in B_{v_1}}$ for all choices of $\sigma_w = \pm 1$ we define the transformation

$$U(\{\boldsymbol{\nu}_{\lambda_w}\}_{w \in B_{v_1}}) \equiv \{\sigma_w \boldsymbol{\nu}_{\lambda_w}\}_{w \in B_{v_1}}, \quad (4.12)$$

and given a set $C \subseteq \mathcal{B}_{v_1}$ we call $\mathcal{U}(C)$ the set of all transformations $U$ such that $\sigma_w = 1$ for $w \notin C$.

If $[2^{h-1}, 2^h)$ is a scale interval $I_h$, $h = 1, 2, \ldots$ we call the first quarter of $I_h$ the *lower part* $I_h^- = [2^{h-1}, \frac{5}{4} 2^{h-1})$ of $I_h$, the fourth quarter of $I_h$ the *upper part* $I_h^+ = [\frac{7}{8} 2^h, 2^h)$ of $I_h$ and the remaining part the *central part* $I_h^c$.

We group the set of branch momenta $\{\boldsymbol{\nu}_{\lambda_w}\}_{w \in \mathcal{B}_{v_1}}$ into *collections* by proceeding iteratively in the way described below. The collections will be built so that in each collection the cancellation discussed in §4.2 above can be exhibited.

Fixed $\boldsymbol{\nu}$ and $h$ choose $\{\boldsymbol{\nu}^1_{\lambda_w}\}_{w \in \mathcal{B}_{v_1}}$ such that $|\boldsymbol{\nu}^1_{v_1}| \in I_h^c$: such $\{\boldsymbol{\nu}^1_{\lambda_w}\}_{w \in \mathcal{B}_{v_1}}$ is called a *representative*. Given the representative we define:

(a) the *branch momenta collection* associated with it to be set of the $\{\boldsymbol{\nu}_{\lambda_w}\}_{w \in \mathcal{B}_{v_1}}$ having the form

$$U(\{\boldsymbol{\nu}^1_{\lambda_w}\}_{w \in \mathcal{B}_{v_1}}), \qquad U \in \mathcal{U}(\mathcal{B}_{1v_1}(\boldsymbol{\nu}, \{\boldsymbol{\nu}^1_{\lambda_w}\}_{w \in \mathcal{B}_{v_1}})), \quad (4.13)$$

and

(b) the *external momenta collection* to be the set of momenta

$$\nu_{v_1}^{1U} = \nu - \sum_{w \in \mathcal{B}_{v_1}} \sigma_w \nu_{\lambda_w}^1, \qquad \text{for } U \in \mathcal{U}(\mathcal{B}_{1v_1}(\nu, \{\nu_{\lambda_w}^1\}_{w \in \mathcal{B}_{v_1}})) \,. \tag{4.14}$$

Note that the elements of the above constructed external momenta collection need not be necessarily contained in $I_h^c$.

We consider then another "representative" $\{\nu_{\lambda_w}^2\}_{w \in \mathcal{B}_{v_1}}$ such that $|\nu_{v_1}^2| \in I_h^c$ *and* not belonging to the branch momenta collection associated with $\{\nu_{\lambda_w}^1\}_{w \in \mathcal{B}_{v_1}}$, if there are any left; and we consider the corresponding branch momenta and external momenta collections as above. We proceed in this way until all the representatives such that $\nu_1$ is in $I_h^c$, for the given $h$, have been put into some collection of branch momenta.

We then repeat the above construction with the interval $I_h^-$ replacing the $I_h^c$, always being careful not to consider representatives $\{\nu_{\lambda_w}\}_{w \in \mathcal{B}_{v_1}}$ that appeared as members of previously constructed collections. It is worth pointing out that not all the external momenta $\nu_{v_1}^U$, $U \in \mathcal{U}(\mathcal{B}_{1v_1}(\nu, \{\nu_{\lambda_w}^1\}_{w \in \mathcal{B}_{v_1}}))$, are in $I_h^-$, but they are all in the corridor $I_{h-1}^+ \cup I_h^-$, by (4.11).

Finally we consider the interval $I_{h-1}^+$, (if $h = 1$ we simply skip this step). The construction is repeated for such intervals.

Proceeding iteratively in this way starting from $h = 1$ and, after exhausting all the $h = 1$ cases, continuing with the $h = 2, 3 \ldots$ cases, we shall have grouped the sets of branch momenta into collections obtainable from a representative $\{\nu_{\lambda_w}\}_{w \in \mathcal{B}_{v_1}}$ by applying the operations $U \in \mathcal{U}(\mathcal{B}_{1v_1}(\nu, \{\nu_{\lambda_w}^1\}_{w \in \mathcal{B}_{v_1}}))$ to it. *Note that, in this way, when the interval $I_{h-1}^+$ is considered, all the remaining representatives are such that $|\nu_{v_1}^U| \in I_{h-1}^+$ for all $U \in \mathcal{U}(\mathcal{B}_{1v_1}(\nu, \{\nu_{\lambda_w}^1\}_{w \in \mathcal{B}_{v_1}}))$*.

REMARK. Note that the graphs with momenta in each collection are just the graphs involved in the parity cancellation described in the previous section. In fact if $U$ is generated by the signs $\{\sigma_w\}_{w \in \mathcal{B}_v}$, we have

$$\nu_{v_1}^U = \Big( \prod_{w \in \mathcal{B}_{1v_1}} U_{v_1 w}^{\sigma_w} \{\nu_x\} \Big)_{v_1}, \qquad (U(\{\nu_{\lambda_{\tilde{w}}}\}_{\tilde{w} \in \mathcal{B}_{v_1}}))_w = \sum_{z \leq w} \Big( \prod_{\tilde{w} \in \mathcal{B}_{1v_1}} U_{v_1 \tilde{w}}^{\sigma_{\tilde{w}}} \{\nu_x\} \Big)_z, \tag{4.15}$$

where, given the sets $\{\nu_x\}$ and $\{\nu_{\lambda_{\tilde{w}}}\}$, $(\{\nu_x\})_v$ denotes the external momentum in $\{\nu_x\}$ corresponding to the node $v$ and $(\{\nu_{\lambda_{\tilde{w}}}\})_w$ denotes the branch momentum in $\{\nu_{\lambda_{\tilde{w}}}\}$ corresponding to the branch $\lambda_w$. Moreover the complexity of the above construction is due to the necessity of avoiding overcountings. In fact it is possible that, for some $U \in \mathcal{U}(\mathcal{B}_{1v_1}(\nu, \{\nu_{\lambda_w}\}_{w \in \mathcal{B}_{v_1}}))$, one has

$$\mathcal{B}_{1v_1}(\nu, U(\{\nu_{\lambda_w}\}_{w \in \mathcal{B}_{v_1}})) \neq \mathcal{B}_{1v_1}(\nu, \{\nu_{\lambda_w}\}_{w \in \mathcal{B}_{v_1}}) \,, \tag{4.16}$$

because the scale of $\nu_{v_1}^U$ may be $h - 1$, while that of $\nu_{v_1}$ may be $h$; so that if one considered, for instance, $I_{h-1}^+$ before $I_h^-$ overcountings would be possible, and in fact they would occurr.

A convenient way to rewrite (4.10) is the following:

$$\sum_\nu |\nu|^s \Big| \sum_{h_{v_1}} \sum_{\{\nu_{\lambda_w}\}_{w \in \mathcal{B}_{v_1}}}^* \sum_{U \in \mathcal{U}(\mathcal{B}_{1v_1})} (i\nu_{v_1}^U)^{\mathbf{a}_{v_1}} f_{\nu_{v_1}^U} \prod_{w \in \mathcal{B}_{v_1}} W(\vartheta_{v_1 w}^0, \sigma_w \nu_{\lambda_w}) \Big| \,, \tag{4.17}$$

where $\sum^*_{\{\nu_{\lambda_w}\}_{w \in \mathcal{B}_{v_1}}}$ means sum over the above defined representatives such that $\nu_{v_1}$ is compatible with $h_{v_1}$; and we abridge $\mathcal{B}_{1v_1}(\nu, \{\nu_{\lambda_w}\}_{w \in \mathcal{B}_{v_1}})$ by $\mathcal{B}_{1v_1}$ in conformity with the notations introduced after (4.11) ($\mathcal{B}_{1v_1}$ is contained in $\mathcal{B}_{v_1}$).

The explicit sum over the scales $h_{v_1}$ is introduced to simplify the bounds analysis that we perform later, see §4.5. Note that $\nu^U_{v_1}$ is, in general, not compatible with $h_{v_1}$, *i.e.* we are grouping together also terms with different scale label (but the difference in scale is at most one).

Noting that by the parity properties of $f_\nu$:

$$W(\vartheta^0_{v_1 w}, \sigma_w \nu_{\lambda_w}) = \sigma_w W(\vartheta^0_{v_1 w}, \nu_{\lambda_w}) \tag{4.18}$$

we have, *still abridging* $\mathcal{B}_{1v_1}(\nu, \{\nu\}_{w \in \mathcal{B}_{v_1}})$ *by* $\mathcal{B}_{1v_1}$, from (4.17):

$$\sum_\nu |\nu|^s \Big| \sum_{h_{v_1}} \sum^*_{\{\nu_{\lambda_w}\}_{w \in \mathcal{B}_{v_1}}} \sum_{U \in \mathcal{U}(\mathcal{B}_{1v_1})} \Big[ \Big(\prod_{w \in \mathcal{B}_{1v_1}} \sigma_w \Big) (i\nu^U_{v_1})^{\mathbf{a}_{v_1}} f_{\nu^U_{v_1}} \Big] \prod_{w \in \mathcal{B}_{v_1}} W(\vartheta^0_{v_1 w}, \nu_{\lambda_w}) \Big| . \tag{4.19}$$

We can apply the interpolation in (4.8) to the node $v$ and rewrite (4.19) as:

$$\sum_\nu |\nu|^s \Big| \sum_{h_{v_1}} \sum^*_{\{\nu_{\lambda_w}\}_{w \in \mathcal{B}_{v_1}}} \Big\{ \sum_{||\mathbf{a}_{v_1}||=p_{v_1}+1} \Big( \prod_{w \in \mathcal{B}_{1v_1}} \int^0_1 dt_w \Big)$$
$$\cdot \Big( \prod_{w \in \mathcal{B}_{1v_1}} \frac{\partial}{\partial t_w} \Big) \Big[ (i\nu_{v_1}(\mathbf{t}_{v_1}))^{\mathbf{a}_{v_1}} f_{\nu_{v_1}(\mathbf{t}_{v_1})} \Big] \Big\} \prod_{w \in \mathcal{B}_{v_1}} W(\vartheta^0_{v_1 w}, \nu_{\lambda_w}) \Big| , \tag{4.20}$$

where if $\mathcal{B}_{1v_1} = \emptyset$ no interpolation is made; and we note that by (4.9), by the definition of nodes out of order and by the iterative grouping of the representatives:

$$2^{h_{v_1}-2} \leq |\nu_{v_1}(\mathbf{t}_{v_1})| < 2^{h_{v_1}} , \tag{4.21}$$

so that the interpolation formulae discussed in Sec. 4.2 *can be used* because no singularity arises in performing the $\mathbf{t}_{v_1}$-integrations.

By the definition of $W(\vartheta^0, \nu)$, and still abridging $\mathcal{B}_{1v_1}(\nu, \{\nu_{\lambda_w}\}_{w \in \mathcal{B}_{v_1}})$ by $\mathcal{B}_{1v_1}$, we can write (4.20) as

$$\sum_\nu |\nu|^s \Big| \sum_{h_{v_1}} \sum^*_{\{\nu_{\lambda_w}\}_{w \in \mathcal{B}_{v_1}}} \Big\{ \sum_{||\mathbf{a}_{v_1}||=p_{v_1}+1} \Big( \prod_{w \in \mathcal{B}_{1v_1}} \int^0_1 dt_w \Big) \cdot$$
$$\Big( \prod_{w \in \mathcal{B}_{1v_1}} \frac{\partial}{\partial t_w} \Big) \Big[ (i\nu_{v_1}(\mathbf{t}_{v_1}))^{\mathbf{a}_{v_1}} f_{\nu_{v_1}(\mathbf{t}_{v_1})} \Big] \Big\} \tag{4.22}$$
$$\prod_{w \in \mathcal{B}_{v_1}} \sum_{[\{\nu_x\}_{x \leq w}; \nu_{\lambda_w}]} \mathrm{Val}(\vartheta^0_{v_1 w}, \{\nu_x\}_{x \leq w}) \Big| ,$$

where the sum over $[\{\nu_x\}_{x \leq w}; \nu_{\lambda_w}]$ is a sum over the $\{\nu_x\}_{x \leq w}$ with $\sum_{x \leq w} \nu_x = \nu_{\lambda_w}$.

If we use (see (4.9)):

$$\frac{\partial}{\partial t_w} \equiv \Big( 2\nu_{\lambda_w} \cdot \frac{\partial}{\partial \nu} \Big)_{\nu = \nu_v(\mathbf{t}_v)} \equiv \Big( \sum_{z \leq w} 2\nu_z \cdot \frac{\partial}{\partial \nu} \Big)_{\nu = \nu_v(\mathbf{t}_v)} , \tag{4.23}$$

to compute differentiations with respect to $t_w$, we can write (4.20) as

$$\sum_{\boldsymbol{\nu}} |\boldsymbol{\nu}|^s \Big| \sum_{h_{v_1}} \sum_{\{\boldsymbol{\nu}_{\lambda_w}\}_{w\in\mathcal{B}_{v_1}}}^{*} \Big\{ \sum_{\|\mathbf{a}_{v_1}\|=p_{v_1}+1} \Big( \prod_{w\in\mathcal{B}_{1v_1}} \int_1^0 dt_w \Big) \cdot$$
$$\cdot \Big( |\boldsymbol{\nu}|^{-(p_{v_1}-q_{v_1})} \frac{\partial^{q_{v_1}}}{\partial \boldsymbol{\nu}^{q_{v_1}}} (i\boldsymbol{\nu})^{\mathbf{a}_{v_1}} f_{\boldsymbol{\nu}} \Big)_{\boldsymbol{\nu}=\boldsymbol{\nu}_{v_1}(\mathbf{t}_{v_1})} \Big\} \cdot$$
$$\cdot \Big[ \prod_{w\in\mathcal{B}_{1v_1}} \Big( \sum_{z\leq w} 2\boldsymbol{\nu}_z \Big) \Big] \Big[ \prod_{w\in\mathcal{B}_{v_1}\setminus\mathcal{B}_{1v_1}} |\boldsymbol{\nu}_{v_1}(\mathbf{t}_{v_1})| \Big]$$
$$\prod_{w\in\mathcal{B}_{v_1}} \Big[ \sum_{[\{\boldsymbol{\nu}_x\}_{x\leq w};\boldsymbol{\nu}_{\lambda_w}]} \mathrm{Val}(\vartheta_{v_1 w}, \{\boldsymbol{\nu}_x\}_{x\leq w}) \Big] \Big|, \qquad (4.24)$$

where we recall that $q_{v_1} = |\mathcal{B}_{1v_1}|$ and the sum over $[\{\boldsymbol{\nu}_x\}_{x\leq w}; \boldsymbol{\nu}_{\lambda_w}]$ again denotes sum over the $\{\boldsymbol{\nu}_x\}_{x\leq w}$ with $\sum_{x\leq w} \boldsymbol{\nu}_x = \boldsymbol{\nu}_{\lambda_w}$; here the factor $|\boldsymbol{\nu}|^{-(p_{v_1}-q_{v_1})}$ (which, computed for $\boldsymbol{\nu} = \boldsymbol{\nu}_{v_1}(\mathbf{t}_{v_1})$, is identical to the inverse of $\prod_{w\in\mathcal{B}_{v_1}\setminus\mathcal{B}_{1v_1}} |\boldsymbol{\nu}_{v_1}(\mathbf{t}_{v_1})|$) has been introduced so that a dimensional estimate of the factor in the second line of (4.24) can be taken proportional to $2^{h_{v_1}(1-b)}$.

If $w \in \mathcal{B}_{v_1} \setminus \mathcal{B}_{1v_1}$ we have:

$$\boldsymbol{\nu}_{v_1}(\mathbf{t}_{v_1}) = X_{v_1 w}(\mathbf{t}_{v_1}) p_{v_1} 2^o \boldsymbol{\nu}_{\lambda_w} \equiv X_{v_1 w}(\mathbf{t}_{v_1}) p_{v_1} 2^o \sum_{z\leq w} \boldsymbol{\nu}_z , \qquad (4.25)$$

where $X_{v_1 w}(\mathbf{t}_{v_1})$ is a matrix with $\|X_{v_1 w}(\mathbf{t}_{v_1})\| \leq \ell^2$, if $\|\cdot\|$ is defined as the sum of the absolute values of the matrix elements. Hence in (4.24) we can write

$$\prod_{w\in\mathcal{B}_{v_1}\setminus\mathcal{B}_{1v_1}} |\boldsymbol{\nu}_{v_1}(\mathbf{t}_{v_1})| = \prod_{w\in\mathcal{B}_{v_1}\setminus\mathcal{B}_{1v_1}} (2^o p_{v_1}) \tilde{\mathbf{x}}_{v_1 w}(\mathbf{t}_{v_1}) \cdot \boldsymbol{\nu}_{\lambda_w}$$
$$= \prod_{w\in\mathcal{B}_{v_1}\setminus\mathcal{B}_{1v_1}} (2^o p_{v_1}) \tilde{\mathbf{x}}_{v_1 w}(\mathbf{t}_{v_1}) \cdot \sum_{z\leq w} \boldsymbol{\nu}_z , \qquad (4.26)$$

where $\tilde{\mathbf{x}}_{v_1 w}(\mathbf{t}_{v_1})$ is a suitable vector depending on $\boldsymbol{\nu}_{\lambda_w}$ *but not on the individual terms* $\boldsymbol{\nu}_z$, and such that $|\tilde{\mathbf{x}}_{vw}(\mathbf{t}_{v_1})| < 1$. The possibility of writing the absolute value of $\boldsymbol{\nu}_{v_1}(\mathbf{t}_{v_1})$ as a sum of the values of the $\boldsymbol{\nu}_z$'s with $\boldsymbol{\nu}_z$–"independent" coefficients, *i.e. the possibility of "linearizing" the absolute value* (in the sense of (4.26)), will be crucial in the following. *It will allow us to treat symmetrically the nodes out of order and the ones that are not out of order.*

We obtain, with the above notations (and taking $o = 5$, see (4.11)):

$$S_k(\vartheta^0) = \sum_{\boldsymbol{\nu}} |\boldsymbol{\nu}|^s \Big| \sum_{h_{v_1}} \sum_{\{\boldsymbol{\nu}_{\lambda_w}\}_{w\in\mathcal{B}_{v_1}}}^{*} \Big\{ \sum_{\|\mathbf{a}_{v_1}\|=p_{v_1}+1} \Big( \prod_{w\in\mathcal{B}_{1v_1}} \int_1^0 dt_w \Big) \cdot$$
$$\cdot \Big( |\boldsymbol{\nu}|^{-(p_{v_1}-q_{v_1})} \frac{\partial^{q_{v_1}}}{\partial \boldsymbol{\nu}^{q_{v_1}}} (i\boldsymbol{\nu})^{\mathbf{a}_{v_1}} f_{\boldsymbol{\nu}} \Big)_{\boldsymbol{\nu}=\boldsymbol{\nu}_{v_1}(\mathbf{t}_{v_1})} \Big\} \cdot$$
$$\Big[ \prod_{w\in\mathcal{B}_{1v_1}} \Big( \sum_{z\leq w} 2\boldsymbol{\nu}_z \Big) \Big] \Big[ \prod_{w\in\mathcal{B}_{v_1}\setminus\mathcal{B}_{1v_1}} \Big( 2^4 p_{v_1} \tilde{\mathbf{x}}_{v_1 w}(\mathbf{t}_{v_1}) \cdot \sum_{z\leq w} 2\boldsymbol{\nu}_z \Big) \Big] \qquad (4.27)$$
$$\prod_{w\in\mathcal{B}_{v_1}} \sum_{\{\boldsymbol{\nu}_x\}_{x\leq w};\boldsymbol{\nu}_{\lambda_w}} \mathrm{Val}(\vartheta_{v_1 w}, \{\boldsymbol{\nu}_x\}_{x\leq w}) \Big|,$$

which we can rewrite by collecting the terms in the following way

$$S_k(\vartheta^0) = \sum_{\boldsymbol{\nu}} |\boldsymbol{\nu}|^s \Big| \sum_{h_{v_1}} \sum_{\{\boldsymbol{\nu}_{\lambda_w}\}_{w \in \mathcal{B}_{v_1}}}^{*} \Big\{ \sum_{||\mathbf{a}_{v_1}||=p_{v_1}+1} \Big( \prod_{w \in \mathcal{B}_{1v_1}} \int_{1}^{0} dt_w \Big) \cdot$$
$$\cdot \Big( \frac{Y_{v_1}(\mathbf{t}_{v_1})}{|\boldsymbol{\nu}|^{p_{v_1}-q_{v_1}}} \frac{\partial^{|\mathcal{B}_{1v_1}|}}{\partial \boldsymbol{\nu}^{|\mathcal{B}_{1v_1}|}} (i\boldsymbol{\nu})^{\mathbf{a}_{v_1}} f_{\boldsymbol{\nu}} \Big)_{\boldsymbol{\nu} = \boldsymbol{\nu}_{v_1}(\mathbf{t}_{v_1})} \Big\} \cdot \qquad (4.28)$$
$$\Big[ \prod_{w \in \mathcal{B}_{v_1}} \Big( \sum_{z \leq w} 2\boldsymbol{\nu}_z \Big) \Big] \prod_{w \in \mathcal{B}_{v_1}} \sum_{\{\boldsymbol{\nu}_x\}_{x \leq w}; \boldsymbol{\nu}_{\lambda_w}} \text{Val}(\vartheta_w, \{\boldsymbol{\nu}_x\}_{x \leq w}) \Big|,$$

where the tensor

$$Y_{v_1}(\mathbf{t}_{v_1}) = \Big[ \prod_{w \in \mathcal{B}_{v_1} \setminus \mathcal{B}_{1v_1}} 2^4 p_{v_1} \tilde{\mathbf{x}}_{v_1 w}(\mathbf{t}_{v_1}) \Big] \qquad (4.29)$$

depends also on $\boldsymbol{\nu}$ and $\{\boldsymbol{\nu}_{\lambda_w}\}_{w \in \mathcal{B}_{v_1}}$, (although this dependence is not shown, to simplify the notation), and has to be contracted with the external momenta $\boldsymbol{\nu}_z$, $z \leq w \in \mathcal{B}_{1v_1}$.

The representation (4.28) for $S_k(\vartheta^0)$ in (4.10) has a form well suited for a repetition of the construction, which will now starts from nodes lower than $v_1$. The repetition of the construction leads eventually to a representation of the values $W(\vartheta_0, \boldsymbol{\nu})$ as a summation over *paths* on $\vartheta_0$: the description of the path representation is the content of the next subsection. The representation will take into full account the cancellations and once achieved the bounds that we are trying to establish will be trivial (in the same sense as the "failed" bounds in §4.1 above).

**4.4. Development of the interpolation; path expansion.** Developing the sum $\sum_{z \leq w} 2\boldsymbol{\nu}_z$ in (4.28) $S_k(\vartheta^0)$ is given by a sum of terms corresponding to a collection of nodes lying on the paths $P(v_1, z(v_1, w))$ leading from $v_1$ to a node $z$: the collection is defined by the "choices" of one particular addend $2\boldsymbol{\nu}_z$ in the sum $\sum_{z \leq w} 2\boldsymbol{\nu}_z$, with $z = z(v_1, w)$, $w \in \mathcal{B}_{v_1}$. Therefore, in general, we can think that (4.28) corresponds to a sum over a collection of paths $P(v_1, z(v_1, w))$ for the $w \in \mathcal{B}_{v_1}$. The paths are regarded as totally ordered (and gapless) sequences of nodes on $\vartheta_0$.

We can call $\mathcal{P}_1$ the family of the possible collections of paths that arise when expanding the sums $\sum_{z \leq w}$ in (4.28): each element $\mathbf{P}_1$ of $\mathcal{P}_1$ can be identified with one contribution to (4.28). And, by using the notation in (4.9) $\mathbf{t}_v = \{t_w\}_{w \in \mathcal{B}_{1v}}$, the result is the following more explicit interpolation formula reexpressing the r.h.s. of (4.28):

$$\sum_{\boldsymbol{\nu}} |\boldsymbol{\nu}|^s \Big| \sum_{h_{v_1}} \sum_{\mathbf{P}_1 \in \mathcal{P}_1} \sum_{\{\boldsymbol{\nu}_{\lambda_w}\}_{w \in \mathcal{B}_{v_1}}}^{*} \Big\{ \sum_{||\mathbf{a}_{v_1}||=p_{v_1}+1} \Big( \prod_{w \in \mathcal{B}_{1v_1}} \int_{1}^{0} dt_w \Big) \cdot$$
$$\cdot \Big( \frac{Y_{v_1}(\mathbf{t}_{v_1})}{|\boldsymbol{\nu}|^{p_{v_1}-q_{v_1}}} \frac{\partial^{q_{v_1}}}{\partial \boldsymbol{\nu}^{q_{v_1}}} (i\boldsymbol{\nu})^{\mathbf{a}_{v_1}} f_{\boldsymbol{\nu}} \Big)_{\boldsymbol{\nu} = \boldsymbol{\nu}_{v_1}(\mathbf{t}_{v_1})} \Big\} \cdot \qquad (4.30)$$
$$\cdot \Big( \prod_{z: P(v_1, z) \in \mathbf{P}_1} 2\boldsymbol{\nu}_z \Big) \cdot \Big( \prod_{w \in \mathcal{B}_{v_1}} \sum_{\{\boldsymbol{\nu}_x\}_{x \leq w}; \boldsymbol{\nu}_{\lambda_w}} \text{Val}(\vartheta^0_{v_1 w}, \{\boldsymbol{\nu}_x\}_{x \leq w}) \Big) \Big|,$$

where the interpolation is considered when $\mathcal{B}_{1v_1} \neq \emptyset$ (*i.e.* when it makes sense), and the indices have to be contracted suitably, and we recall that $q_{v_1} \equiv |\mathcal{B}_{1v_1}|$.

The above formula can be rewritten as:

$$\sum_{\nu} |\nu|^s \Big| \sum_{h_{v_1}} \sum_{\{\nu_{\lambda_w}\}_{w \in \mathcal{B}_{v_1}}}^{*} \sum_{\mathbf{P}_1 \in \mathcal{P}_1} \Big( \prod_{v \in [\mathbf{P}_1]} \sum_{\{\nu_{\lambda_y}\}_{y \in \mathcal{B}_v}} \Big) $$
$$\Big\{ \sum_{||\mathbf{a}_{v_1}||=p_{v_1}+1} \Big( \prod_{w \in \mathcal{B}_{1v_1}} \int_1^0 dt_w \Big) \cdot \Big( \frac{Y_{v_1}(\mathbf{t}_{v_1})}{|\nu|^{p_{v_1}-q_{v_1}}} \frac{\partial^{q_{v_1}}}{\partial \nu^{q_{v_1}}} (i\nu)^{\mathbf{a}_{v_1}} f_\nu \Big)_{\nu=\nu_{v_1}(\mathbf{t}_{v_1})} \Big\} \qquad (4.31)$$
$$\cdot \prod_{v \in [\mathbf{P}_1]} \sum_{||\mathbf{a}_v||=p_v+1} (2\nu_v)^{\eta_v} (i\nu_v)^{\mathbf{a}_v+1} f_{\nu_v} \prod_{y \in \mathcal{B}_v/[\mathbf{P}_1]} W(\vartheta^0_{vy}, \nu_{\lambda_y}) \Big| ,$$

where $[\mathbf{P}_1] = \bigcup_{w \in \mathcal{B}_{v_1}} P(v_1, z(v_1, w))/\{v_1\}$ and $\eta_v$ is equal to 1 if $v = z(v_1, w)$ for some $w \in \mathcal{B}_{v_1}$ and 0 otherwise.

We are now in position to iterate the resummation done in the previous section leading from (4.10) to (4.17) and "concerning" the highest node $v_1$. For each $\tilde{v} \in \mathbf{P}_1$, $\tilde{v} < v_1$, let $h_{\tilde{v}} = h_{\tilde{v}}(\nu_{\lambda_{\tilde{v}}}, \{\nu_{\lambda_w}\}_{w \in \mathcal{B}_{\tilde{v}}})$ be *the scale of* $\nu_{\tilde{v}}$, i.e. $\nu_{\tilde{v}} = \nu_{\lambda_{\tilde{v}}} - \sum_{w \in \mathcal{B}_{\tilde{v}}} \nu_{\lambda_w}$ is such that $2^{h_{\tilde{v}}-1} \leq |\nu_{\tilde{v}}| < 2^{h_{\tilde{v}}}$.

Given an immediate predecessor $w$ of $\tilde{v}$ we say that $w$ is *out of order* with respect to $\tilde{v}$ if

$$2^{h_{\tilde{v}}} > 2^5 p_{\tilde{v}} |\nu_{\lambda_w}| , \qquad (4.32)$$

where $p_{\tilde{v}}$ is the number of branches entering $\tilde{v}$. We denote $\mathcal{B}_{1\tilde{v}} \equiv \mathcal{B}_{1\tilde{v}}(\nu_{\lambda_{\tilde{v}}}, \{\nu_{\lambda_w}\}_{w \in \mathcal{B}_{\tilde{v}}}) \subseteq \mathcal{B}_{\tilde{v}}$ the nodes $w \in \mathcal{B}_{\tilde{v}}$ which are out of order with respect to $\tilde{v}$.

Given a set $\{\nu_{\lambda_w}\}_{w \in B_{\tilde{v}}}$ for all choices of $\sigma_w = \pm 1$ we define

$$U(\{\nu_{\lambda_w}\}_{w \in B_{\tilde{v}}}) \equiv \{\sigma_w \nu_{\lambda_w}\}_{w \in B_{\tilde{v}}} , \qquad (4.33)$$

and given a set $C \subseteq \mathcal{B}_{\tilde{v}}$ we call $\mathcal{U}(C)$ the set of all transformations such that $\sigma_w = 1$ for $w \notin C$.

We group the set of branch momenta $\{\nu_{\lambda_w}\}_{w \in \mathcal{B}_{\tilde{v}}}$ and the external momenta into *collections* by proceeding, very closely following the preceding construction, with $\nu_{\lambda_w}$ playing the role of $\nu$, in the way described below.

Fixed $\nu_{\lambda_{\tilde{v}}}$ and $h$ we choose a $\{\nu^1_{\lambda_w}\}_{w \in \mathcal{B}_{\tilde{v}}}$ such that $|\nu^1_{\tilde{v}}| \in I^c_h$ where $\nu^1_{\tilde{v}} = \nu_{\lambda_{\tilde{v}}} - \sum_{w \in \mathcal{B}_{\tilde{v}}} \nu_{\lambda_w}$.

Then $\{\nu^1_{\lambda_w}\}_{w \in \mathcal{B}_{\tilde{v}}}$ is called a *representative*. For such representative we define the *branch momenta collection*, associated with it to be the set of the $\{\nu_{\lambda_w}\}_{w \in \mathcal{B}_{\tilde{v}}}$ having the form $U(\{\nu^1_{\lambda_w}\}_{w \in \mathcal{B}_{\tilde{v}}})$ and the *external momenta collection* to be the set of momenta $\nu^{1U}_{\tilde{v}} = \nu - \sum_{w \in \mathcal{B}_{\tilde{v}}} \sigma_w \nu^1_{\lambda_w}$, for $U \in \mathcal{U}(\mathcal{B}_{1\tilde{v}}(\nu_{\lambda_{\tilde{v}}}, \{\nu_{\lambda_w}\}_{w \in \mathcal{B}_{\tilde{v}}})/[\mathbf{P}_1])$. Note again that the above constructed external momenta collection is not necessarily contained in $I^c_h$.

We consider then another "representative" $\{\nu^2_{\lambda_w}\}_{w \in \mathcal{B}_{\tilde{v}}}$ such that $|\nu^2_{\tilde{v}}| \in I^c_h$ and does not belong to the just constructed branch momenta collection associated with $\{\nu^1_{\lambda_w}\}_{w \in \mathcal{B}_{\tilde{v}}}$, if there is any; and then we consider the branch momenta collections and external momenta collections obtained from $\{\nu^2_{\lambda_w}\}_{w \in \mathcal{B}_{\tilde{v}}}$ by the corresponding $U$ transformations. And, as previously done, we proceed in this way until all the representatives such that $\nu^1_{\tilde{v}}$ is in $I^c_h$ are in some external momenta collections.

The construction is repeated for the interval $I^-_h$, always being careful not to consider $\{\nu_{\lambda_w}\}_{w \in \mathcal{B}_{\tilde{v}}}$ that have been already considered, and finally for the interval $I^+_{h-1}$, see §4.3.

Proceeding iteratively in this way and considering the same sequence of $h$'s as in the previous case (*i.e.* the natural $h = 1, 2, \ldots$), at the end we shall have grouped the set of branch momenta into collections obtainable from a "representative" $\{\nu_{\lambda_w}\}_{w \in \mathcal{B}_{\tilde{v}}}$ by applying the operations $U \in \mathcal{U}(\mathcal{B}_{1\tilde{v}}(\nu_{\lambda_w}, \{\nu^1_{\lambda_w}\}_{w \in \mathcal{B}_{\tilde{v}}}) \setminus [\mathbf{P}_1])$ to it.

*In other words the definition of the representatives* $\{\boldsymbol{\nu}_{\lambda_w}\}_{w \in \mathcal{B}_{\tilde{v}}}$ *is identical to the one for $v_1$ except that the collections are defined only by transformations changing the branch momentum of the lines emerging from the nodes in $\mathcal{B}_{1\tilde{v}}$ but not in $\mathbf{P}_1$.*

We repeat the above construction for all $\tilde{v} \in \mathbf{P}_1$ until all the $\tilde{v} \in \mathbf{P}_1$ are considered starting from the $\tilde{v}$ with $\tilde{v}' = v$ and, after exhausting them, continuing with $\hat{v}$ with $\hat{v}' = \tilde{v}$ and so on. We call $\mathcal{B}_{\tilde{v}}(\mathbf{P}_1)$ the nodes $w$ immediately preceding $\tilde{v}$ *but which are not on the union of the paths* $P \in \mathbf{P}_1$, and $\mathcal{B}_{1\tilde{v}}(\mathbf{P}_1)$ the nodes in $\mathcal{B}_{\tilde{v}}(\mathbf{P}_1)$ which are out of order with respect to $\tilde{v}$; the set of just described transformations will be denoted by $\mathcal{U}(\mathcal{B}_{1\tilde{v}}(\mathbf{P}_1))$.

Proceeding as we did for the highest node $v_1$ and by performing the analogues of the transformations leading from (4.17) to (4.31), we construct for each $\tilde{v} \in \mathbf{P}_1$ new paths $\mathbf{P}_2$ which, by construction, *will not have common branches* with those in $\mathbf{P}_1$; call $\mathcal{P}_2$ the collection of the pairs $\mathbf{P}_1, \mathbf{P}_2$. The crucial point is that the factors $\tilde{\mathbf{x}}_{vw}(\mathbf{t}_v)$ are the same for all the terms generated by the action of $U \in \mathcal{U}(\mathcal{B}_{1v}(\mathbf{P}_1))$, by (4.26). We iterate then this procedure.

Eventually we end up by constructing a *pavement* $\mathbf{P}$ of the graph with nonoverlapping paths (and the union of the paths does cover the graph); note that the paths are "ordered", (see §3.1), in the sense that they are formed only by comparable lines.

We call $\mathcal{P}$ the collection of all such pavements; $\mathcal{B}_v(\mathbf{P})$, $\mathbf{P} \in \mathcal{P}$, will be the set of nodes $w$ immediately preceding $v$ and such that a path $P(v, z(v, w)) \in \mathbf{P}$ starting from $v$ passes through $w$, and $\mathcal{B}_{1v}(\mathbf{P})$ is the collection of nodes in $\mathcal{B}_v(\mathbf{P})$ out of order with respect to $v$. Note that in general $\mathcal{B}_v(\mathbf{P}) \subseteq \mathcal{B}_v$ (unless $v$ is the highest node $v_1$, when $\mathcal{B}_{v_1}(\mathbf{P}) = \mathcal{B}_{v_1}$).

The set of "path head" nodes $v$, *i.e.* the upper endnodes of paths in $\mathbf{P}$, will be denoted $M_h(\mathbf{P})$: hence if $v \notin M_h(\mathbf{P})$ (*i.e.* if no path in $\mathbf{P}$ has $v$ as path head) then $\mathcal{B}_v(\mathbf{P}) = \emptyset$; likewise $M_e(\mathbf{P})$ will denote the set of "path end" nodes, *i.e.* the nodes $z$ such that $P(v, z)$ is a path in $\mathbf{P}$.

Then we see that (4.30) leads the following *path expansion* for $S_k(\vartheta^0)$ (see (4.10)) summarizing our analysis:

$$S_k(\vartheta^0) = \sum_{\boldsymbol{\nu}} |\boldsymbol{\nu}|^s |W(\vartheta^0, \boldsymbol{\nu})| = \sum_{\boldsymbol{\nu}} |\boldsymbol{\nu}|^s \Big| \sum_{\{h_x\}} \sum_{\mathbf{P} \in \mathcal{P}} \sum_{\{\boldsymbol{\nu}_\lambda\}}^{*} \prod_{v \in M_h(\mathbf{P})} \tag{4.34}$$

$$\Big\{ \Big( \prod_{w \in \mathcal{B}_{1v}(\mathbf{P})} \int_1^0 dt_w \Big) \sum_{||\mathbf{a}_v|| = p_v + 1} \mathcal{O}_v \Big( (i\boldsymbol{\nu}_v(\mathbf{t}_v))^{\mathbf{a}_v} Y_v(\mathbf{t}_v) f_{\boldsymbol{\nu}_v(\mathbf{t}_v)} \Big) \Big\} \Big|,$$

where the sum over $\{\boldsymbol{\nu}_\lambda\}$ comes with the restriction, among others met in the above derivation, that the external momentum configuration $\{\boldsymbol{\nu}_x\}$ is compatible with the scales $\{h_x\}$; furthermore $\mathcal{O}_v$ is defined as:

$$\mathcal{O}_v \Big( (i\boldsymbol{\nu}_v(\mathbf{t}_v))^{\mathbf{a}_v} Y_v(\mathbf{t}_v) f_{\boldsymbol{\nu}_v(\mathbf{t}_v)} \Big) =$$
$$= \Big( \frac{Y_v(\mathbf{t}_v)}{|\boldsymbol{\nu}|^{|\mathcal{B}_v(\mathbf{P})| - |\mathcal{B}_{1v}(\mathbf{P})|}} \frac{\partial^{|\mathcal{B}_{1v}(\mathbf{P})|}}{\partial \boldsymbol{\nu}^{|\mathcal{B}_{1v}(\mathbf{P})|}} (i\boldsymbol{\nu})^{\mathbf{a}_v} (2\boldsymbol{\nu})^{\eta_v} f_{\boldsymbol{\nu}} \Big)_{\boldsymbol{\nu} = \boldsymbol{\nu}_v(\mathbf{t}_v)}, \tag{4.35}$$

with $Y_v(\mathbf{t}_v)$ defined as

$$Y_v(\mathbf{t}_v) = \begin{cases} \prod_{w \in \mathcal{B}_v(\mathbf{P}) \setminus \mathcal{B}_{1v}(\mathbf{P})} (2^4 p_v \tilde{\mathbf{x}}_{vw}(\mathbf{t}_v)), & \text{if } v \in M_h(\mathbf{P}), \\ 1, & \text{otherwise}, \end{cases} \tag{4.36}$$

and we introduced a label $\eta_v$ associated with each node $v$, such that $\eta_v = 1$ if $v \in M_e(\mathbf{P})$, and $\eta_v = 0$ otherwise.

In the bounds we shall replace the sum over the $\{\boldsymbol{\nu}_\lambda\}$ by a sum over $\{\boldsymbol{\nu}_x\}$ compatible with $\{h_x\}$ and forget that only "representatives" have to be counted.

The "paths" $P(v, z)$ can be viewed as "pointers" indicating that an external momentum $\boldsymbol{\nu}_v$ has been "shifted" to a node $z$ becoming $\boldsymbol{\nu}_z$ in the sense of the remark following (4.9). Therefore it will not be surprising that the bounds derived in §4.5 below give the result described in the remark. One can see that the path representation is in some sense playing the role of an integration by parts and "it shifts the derivatives" (*i.e.* the factors $\boldsymbol{\nu}$'s, as we work on Fourier representations) where they are less "dangerous". The cancellations simply tell us that this feat can be achieved although *no integration* is performed on our expressions.

**4.5. Bounds.** If we show that the sum in the l.h.s. of (4.34) is bounded by $C_1^k \prod_v p_v!$, for some $C_1 > 0$ then we shall have proved that $\mathbf{h}$ is analytic for $\varepsilon < C_1^{-1}$ and of class $C^{(s)}(\mathbb{T}^\ell)$ because the sum over the graphs $\vartheta^0$ is a sum of over $k! \prod_v p_v!^{-1}$ ways of assigning the labels that distinguish for each "shape" of $\vartheta^0$ (recall that in §3 we defined the graphs to have distinguishable branches and made the otherwise identical branches distinct by attaching to them a label that played no role other than that of making the combinatorics somewhat easier, see §3.1) the branches of $\vartheta^0$ (and the number of "shapes" is $\leq 4^k$). Recall also that there is an overall $\frac{1}{k!}$ that has to divide the l.h.s. of (4.34) to connect it with $\mathbf{h}^{(k)}$, see (3.4).

By the definition of $\eta_v$ after (4.35), when we differentiate with respect to $\frac{\partial^{q_v}}{\partial \boldsymbol{\nu}^{q_v}}$, $q_v = |\mathcal{B}_{1v}(\mathbf{P})|$, the derivatives act on the product of $f_{\boldsymbol{\nu}_v}(\mathbf{t}_v)$ times up to $p_v + 1$ momentum factors, if $v \notin M_e(\mathbf{P})$, or up to $p_v + 2$ factors if $v \in M_e(\mathbf{P})$. Hence we can choose the $q_v \leq p_v$ factors to be differentiated in $\leq \binom{p_v+3}{q_v} \leq 2^{p_v+3}$ ways.

Furthermore the derivatives produce always the same "dimensional bound" as far as the dependence on $h_v$ is concerned. One finds (see also (4.21) and the comments following it, and recall the definition of $\eta_v$ following (4.35)):

$$|\boldsymbol{\nu}|^s \prod_{v \in \vartheta^0} \|\mathcal{O}_v\left((i\boldsymbol{\nu}_v(\mathbf{t}_v))^{\mathbf{a}_v} Y_v(\mathbf{t}_v) f_{\boldsymbol{\nu}_v(\mathbf{t}_v)}\right)\| \leq$$
$$\leq \prod_{v \in \vartheta^0} D_1 D_2^{p_v} q_v! p_v^{p_v-q_v} 2^{h_v(1-b+s+\eta_v)} \leq \prod_{v \in \vartheta^0} D_3 D_4^{p_v} p_v! 2^{h_v(1-b+s+\eta_v)} , \quad (4.37)$$

for suitable constants $D_j$; (4.37) takes into account also a factor $\ell^{p_v+1+\eta_v}$ arising from number of terms in the sum $\sum_{\|\mathbf{a}_v\| \leq p_v+1}$, each of which has been bounded in deriving (4.37) by the maximum over the $\mathbf{a}$'s, and the combinatorial factor from the derivatives. The factor $|\boldsymbol{\nu}|^s$ is here generously bounded by $\prod_v |\boldsymbol{\nu}_v|^s$.

For the purpose of comparison with §4.1, (4.2) and (4.4), by using (4.37) we can bound the sum (4.34) by replacing the sum over $\{h_\lambda\}, \{\boldsymbol{\nu}_\lambda\}$ by a sum over $\{h_x\}$ and over the $\{\boldsymbol{\nu}_x\}$ compatible with $\{h_x\}$ (*i.e.* such that $h_x$ is the scale of $\boldsymbol{\nu}_x$) and by writing

$$\sum_{\boldsymbol{\nu}} |\boldsymbol{\nu}|^s |W(\vartheta^0, \boldsymbol{\nu})| \leq \sum_{\{h_x\}} \sum_{\mathbf{P} \in \mathcal{P}} \sum_{\{\boldsymbol{\nu}_x\} \text{ comp } \{h_x\}} B_2^k \prod_{v \in \vartheta^0} p_v! 2^{h_v(1-b+s+\eta_v)} \leq$$
$$\leq \sum_{\{h_x\}} \sum_{\mathbf{P} \in \mathcal{P}} B_1^k \left( \prod_{v \in \vartheta^0} p_v! 2^{h_v(1-b+s+\ell)} \right) \cdot \prod_{z : P(v,z) \in \mathbf{P}} 2^{h_z} \leq \quad (4.38)$$
$$\leq \sum_{\{h_x\}} \sum_{\mathbf{P} \in \mathcal{P}} B_1^k \prod_{v \in \vartheta^0} p_v! 2^{h_v(1-b+s+\ell+p_v)} \prod_{z : P(v,z) \in \mathbf{P}} 2^{h_z - h_v} ,$$

for suitable constants $B_1 > B_2$, (we use that $\sum_{v \in \vartheta_0} p_v = k - 1$ and that there are $O(2^{h\ell})$ momenta of scale $h$). Note that of all the restrictions that have to be imposed on $\{\boldsymbol{\nu}_x\}$ we only retain here that the scales of $\{\boldsymbol{\nu}_x\}$ are the $\{h_x\}$, *i.e.* $\{\boldsymbol{\nu}_x\}$ comp $\{h_x\}$, (see §4.1).

Hence, by applying (4.2)÷(4.4), one finds that the sum (4.10), once reexpressed as (4.34) and remembering that $b = 2 + s + l + \mu$, is bounded above by:

$$\sum_{\boldsymbol{\nu}} |\boldsymbol{\nu}|^s |W(\vartheta^0, \boldsymbol{\nu})| \leq \sum_{\{h_x\}} \sum_{\mathbf{P} \in \mathcal{P}} B_1^k \Big( \prod_{v \in \vartheta^0} p_v! \Big) \cdot \Big( \prod_{P(v,z) \in \mathbf{P}} 2^{h_z - h_v} \Big) \cdot$$
$$\cdot \Big[ 2^{-h_{v_1}} \Big( \prod_{v < v_1} 2^{-(h_v - h_{v'})} \Big) \Big( \prod_{v \leq v_1} 2^{-h_v \mu} \Big) \Big] = \qquad (4.39)$$
$$= \sum_{\{h_x\}} \sum_{\mathbf{P} \in \mathcal{P}} B_1^k \Big( \prod_{v \in \vartheta} p_v! \Big) 2^{-h_{v_1}} \Big( \prod_{v \leq v_1} 2^{-h_v \mu} \Big) \ ,$$

where the second equality is obtained expressing $h_z - h_v$ as the sum of the scale variations along the branches of the path $P(v, z)$ joining $v$ to $z$. The highest node $v_1$ is "privileged" (because of the "extra" $2^{-h_{v_1}}$ factor) since it is the only node for which we are sure that $\eta_v = 0$ because $v_1 \notin M_e(\mathbf{P})$.

Of course in deriving (4.39) one has to take into account *also* the cut off factors $e^{-\kappa |\boldsymbol{\nu}|}$ appearing in the Fourier coefficients $f_{\boldsymbol{\nu}_v}$, which may be stricken by differentiations: but their contribution is not worse than the terms that one would obtain if they were absent. In fact they generate (when differentiated with respect to the interpolation paramenters) terms like $\exp(-\kappa 2^{h_v}) \kappa |\boldsymbol{\nu}_z| \leq 2^{-(h_v - h_z)}$. So we can think that their contributions are accounted for in (4.39) by an appropriate choice of $B_1$ and by the $\prod_v p_v!$ which may pop up also when treating such terms (via $e^{-\kappa 2^{h_v}} \kappa^{p_v} \leq p_v! 2^{-p_v h_v}$).

*It is now clear that (4.39) is a much better bound than the "naive" bound (4.4).* In fact using that the number of pavement $\mathbf{P}$ in $\mathcal{P}$ is bounded by $2^k$, see Appendix A2, we get:

$$\sum_{\boldsymbol{\nu}} |\boldsymbol{\nu}^s| \Big| W(\vartheta_0, \boldsymbol{\nu}) \Big| \leq B^k \Big( \prod_{v \in \vartheta_0} p_v! \Big) \sum_{\{h_x\}} \Big[ 2^{-h_v(1+\mu)} \prod_{v < v} 2^{-\mu h_v} \Big] \qquad (4.40)$$

for some $B > 0$.

REMARK. Comparing (4.40) with (4.4) we see that what has been achieved is to lose completely the favourable factors $2^{h_{v'} - h_v}$ with $h_{v'} \leq h_v$, *which however had no influence on the convergence of (4.4)*, and to gain new factors $2^{-(h_{v'} - h_v)}$ compensating the unfavourable $2^{h_{v'} - h_v}$, $h_{v'} > h_v$, which *did ruin* the convergence of the naive bound (4.4). In the end we have no trace left neither of the "bad" factors in (4.4) nor of the "good" ones: the estimates have become completely marginal, as usual in the interesting problems, and convergence can be decided simply by requiring enough differentiablity on the perturbation $f$, a mild condition on $\mu$, i.e. on $p$ ($p > 2$, see below). This explains also why no resummations are performed for the pairs of nodes $(v, w)$, $v = w'$, both belonging to the same path $P \in \mathbf{P}$: in fact in this way one would obtain more than one factor $2^{h_w - h_v}$, so that one of such factors would still cancel the factor $2^{h_v - h_w}$ appearing in (4.4), but the other "extra" factors would be source of problems for $h_w > h_v$.

The summability over the scale labels follows, for $\mu > 0$. The summability over the orders $k$, weighed by $\varepsilon^k$, follows (as remarked above) by taking into account that the number of graphs with given $p_v$'s (and without node labels) is bounded by $k! / \prod_v p_v!$ (Cayley's formula, see [HP]). Hence Theorem 1.4 is proved, and it follows that if $f \in \hat{C}^{(2+s+\mu)}(\mathbb{T}^\ell)$, $\mu > 0, s \geq 0$, then $\mathbf{h}$ exists and is analytic in $\varepsilon$, for small $\varepsilon$, and it is in $C^{(s)}(\mathbb{T}^\ell)$.

Since the convergence is uniform in the ultraviolet cut off $\kappa$ (which does not appear at all in (4.40)) and since the extra terms that arise in (4.39) due to the presence of $\kappa$ have $\kappa$ as a factor, we see that in the limit $\kappa \to 0$ one can simply disregard such terms. As a consequence of this uniformity the limit as $\kappa \to 0$ of the (4.39) is simply (4.39) itself *without the terms depending on $\kappa$*: this remark completes the description of the cancellations and provides a concrete rule to compute the resummed series.

Thus if $\ell = 1$ we find that for $|\varepsilon|$ small $\mathbf{h}$ is analytic in $\varepsilon$ if $p > 2$ and $\mathbf{h} \in C^{(0)}(\mathbb{T}^1)$. But this result seems to give a regularity weaker than the one found in the exact solution in the example in §2.2. However we have checked that, in that particular example, there are other cancellations and that the above argument performed for that special case would lead to the stronger result (*i.e.* $\mathbf{h} \in C^{(1)}(\mathbb{T}^1)$): hence the condition on $p$ in Theorem 1.4 may be (in some sense) "best possible".

## 5. Multiscale analysis for the hamiltonian model

We come back to the model (1.3), and consider again (for simplicity) a perturbation of the form $f_\nu = |\nu|^{-b}$. We also take $J = 1$, see (1.1), (1.5).

**5.1. Infrared and ultraviolet scales.** Let us define $\chi(x)$ as the characteristic function of the set $\{x \in \mathbb{R} : |x| \in [1/2, 1)\}$, and $\chi_1(x)$ as the characteristic function of the set $\{x \in \mathbb{R}, |x| > 1\}$. Then, if $g_\lambda = [i\boldsymbol{\omega} \cdot \boldsymbol{\nu}_\lambda]^{-2}$ is the propagator (or divisor), see (3.3), we can decompose:

$$g_\lambda = \frac{\chi_1(\boldsymbol{\omega} \cdot \boldsymbol{\nu}_\lambda)}{(i\boldsymbol{\omega} \cdot \boldsymbol{\nu}_\lambda)^2} + \sum_{n=-\infty}^{0} \frac{\chi(2^{-n}\boldsymbol{\omega} \cdot \boldsymbol{\nu}_\lambda)}{(i\boldsymbol{\omega} \cdot \boldsymbol{\nu}_\lambda)^2} = \sum_{n=-\infty}^{1} g_\lambda^{(n)} , \qquad (5.1)$$

*i.e.* we decompose $g_\lambda$ in *infrared* scale components and we introduce the ultraviolet scale labels as in §4.

Inserting the above decompositions in the definition of the value of a graph ((3.3) with $X = 2$), we see that the value of each graph is decomposed into various addends. We can identify the addends simply by attaching to each line $\lambda$ a scale label $n_\lambda \leq 1$, see (5.1).

**5.2. Clusters and resonances.** Besides the new "ultraviolet" cancellations exhibited in §4 we must take into account the old "infrared" cancellations (that are sufficient to prove the existence of Kolmogorov's tori, *i.e.* to solve the problem in the analytic case). Hence we introduce, see [E,G2,GM1], the notion of *clusters* of a graph $\vartheta$.

Given a graph $\vartheta$, a cluster of scale $n \leq 1$ is a maximal set of nodes connected by lines of scale $\geq n$. Sometimes we shall find it useful to think of a cluster as the set of lines connecting the nodes of a cluster. A line $\lambda$ which connects nodes both inside a cluster $T$ is said to be "internal" to the cluster, ($\lambda \in T$), while the lines which connect a node inside with a node outside the cluster are called "external" to the cluster; we say that a line $\lambda$ intersect a cluster $T$, ($\lambda \cap T \neq \emptyset$), if $\lambda$ is internal or external to $T$, *i.e.* if at least one extreme of $\lambda$ is inside $T$. A line is "outside" the cluster $T$ if it is neither internal nor external to it.

The nodes of a cluster $V$ of scale $n_V$ may be linked to other nodes by lines of lower (*i.e.* more negative) scale. Such lines are called "incoming" if they point at a node in the cluster or "outgoing" otherwise (recall, see §3, that the lines are "arrows", *i.e.* are oriented towards the root and establish a partial ordering on the graph regarded as a tree graph). There may be several incoming lines (or zero) but at most one outgoing line, because of the tree structure of the graphs that we consider.

The key notion for the convergence proofs in the analytic case is the notion of *resonance*, or resonant cluster, due to Eliasson, [E]. We use here a version of it, inspired by [GM1].

DEFINITION (RESONANCE). *We define a real resonance a cluster $V$ such that:*
*(1) there is only one incoming line $\lambda_V$ and one outgoing line $\lambda'_V$ and they carry the same momentum;*
*(2) if $n_V$ is the scale of the cluster and $n_{\lambda_V}$ is the scale of the line $\lambda_V$, one has $n_V \geq n_{\lambda_V} + 3$.*
*Likewise we define a virtual resonance a cluster $V$ such that (2) holds and (1) is replaced with:*
*(1') there is only one incoming line $\lambda_V$ and one outgoing line $\lambda'_V$ and they carry opposite momentum.*

It will appear that a resonance definition based on a scale difference $n_V - n_{\lambda_V} = o$ with *any* integer $o \geq 2$ would suffice in order to prove Theorem 1.5; the important feature is that there must be a scale difference *strictly bigger than 1*, in order to allow us to bound the small divisors corresponding to the lines internal to a resonance $V$ with the values they would have by setting equal to zero the momentum flowing through the incoming line, *i.e.* $\boldsymbol{\nu}_{\lambda_V} = \mathbf{0}$, (see below, §5.6 and Appendix A1). The actual value chosen for $o$ may have relevance only for the bounds size or ease.

Then the following result, which is an extension of a known bound due to Siegel and Eliasson, see [S,E], holds.

LEMMA (SIEGEL-ELIASSON'S BOUND). *If we consider only graphs with no real resonances, then there is a constant $C$ such that:*

$$\prod_{\lambda \in \vartheta} \frac{1}{|\boldsymbol{\omega} \cdot \boldsymbol{\nu}_\lambda|} \leq C^k \frac{\prod_{v \in \vartheta} |\boldsymbol{\nu}_v|^{3\tau}}{(\sum_{v \in \vartheta} |\boldsymbol{\nu}_v|)^\tau} . \tag{5.2}$$

In the following the exponent 3 in (5.2) will be denoted $\frac{\eta}{2}$, as this will be useful in §6 and in the concluding remarks. The proof of the above lemma involves a simple extension of the ideas introduced in the well known proofs by Siegel and Eliasson. In Appendix A1 we adapt to our definitions the proof in [E]. The exponent $\eta = 6$ might be not optimal.

**5.3. Superficial renormalization.** Consider a graph $\vartheta$ and call $\hat{\vartheta}$ the graph obtained by deleting the infrared scale labels $\{n_\lambda\}$ and $\vartheta^0$ the graph obtained by deleting the scale and external momentum labels. Therefore $\vartheta$ can be written $\vartheta = (\hat{\vartheta}, \{n_\lambda\})$, or $(\vartheta^0, \{n_\lambda\}, \{\boldsymbol{\nu}_x\})$.

Suppose that the set of scales $\{n_\lambda\}$ is consistent with the existence of a fixed family $\mathbf{V}_1$ of *maximal real resonances*: *i.e.* of real resonances not contained in any larger real resonance. If $V \in \mathbf{V}_1$ we call $\lambda_V = v_b^V v_1^V$ the line incoming into the real resonance and $n_{\lambda_V}$ its scale; likewise $\lambda'_V = v^V v_a^V$ is the outgoing line. Here $v_a^V, v_b^V \in V$ while $v^V, v_1^V$ are out of it (and $v$ could be the root). By definition it must be $n_{\lambda_V} + 3 \leq n_V$.

We consider the graph values at fixed set of scales for the lines not in any $V \in \mathbf{V}_1$ (in particular the scale of the line entering the real resonance is held fixed) and we say that such a set of scales is "compatible" with $\mathbf{V}_1$, denoting this property by $\{n_\lambda\} \& \mathbf{V}_1$.

We introduce the *momentum flowing on* $\lambda_v \in V$ *intrinsic to the cluster* $V$ as $\boldsymbol{\nu}^0_{\lambda_v} = \sum_{v \geq w \in V} \boldsymbol{\nu}_w$, and define the *resonance path* $Q_V$ as the totally ordered path of lines joining the line incoming into the real resonance $V$ and the outgoing line and *not* including the latter two lines. Then the sum over $\{n_\lambda\}$ of the graph values $\text{Val}(\hat{\vartheta}, \{n_\lambda\})$ at fixed $\hat{\vartheta}$, by the definition in (3.1) of the value of a graph, can be written:

$$\sum_{\{n_\lambda\}} \text{Val}(\hat{\vartheta}, \{n_\lambda\}) = (-1)^k \sum_{\mathbf{V}_1} \sum_{\{n_\lambda\} \& \mathbf{V}_1} \left(\prod_{v \in \vartheta} f_{\boldsymbol{\nu}_v}\right) \cdot \left[\prod_{\substack{\lambda \cap \mathbf{V}_1 = \emptyset \\ \lambda = xy}} \frac{\boldsymbol{\nu}_x \cdot \boldsymbol{\nu}_y \, \chi(2^{-n_\lambda} \boldsymbol{\omega} \cdot \boldsymbol{\nu}_\lambda)}{(\boldsymbol{\omega} \cdot \boldsymbol{\nu}_\lambda)^2}\right].$$

$$\cdot \left\{ \prod_{V \in \mathbf{V}_1} \left[ \frac{\boldsymbol{\nu}_0 \cdot \boldsymbol{\nu}_a \, \boldsymbol{\nu}_1 \cdot \boldsymbol{\nu}_b \, \chi(2^{-n_\lambda} \boldsymbol{\omega} \cdot \boldsymbol{\nu}_{\lambda_V})}{(\boldsymbol{\omega} \cdot \boldsymbol{\nu}_{\lambda_V})^4} \right] \mathcal{V}(\boldsymbol{\omega} \cdot \boldsymbol{\nu}_{\lambda_V} | V, \{n_\lambda\}_{\lambda \in V}) \right\}, \quad (5.3)$$

where a line $\lambda$ is defined to intersect a real resonance $V \in \mathbf{V}_1$ if it is internal to $V$, or $\lambda \in \{\lambda_V, \lambda'_V\}$ (*i.e.* $\lambda$ is not outside $V$ in the sense of §5.2), and the *resonance value* $\mathcal{V}$ is defined by:

$$\mathcal{V}(\zeta | V, \{n_\lambda\}_{\lambda \in V}) = \prod_{\substack{\lambda \in V \\ \lambda = xy}} \boldsymbol{\nu}_x \cdot \boldsymbol{\nu}_y \frac{\chi(2^{-n_\lambda}(\boldsymbol{\omega} \cdot \boldsymbol{\nu}_\lambda^0 + \sigma_\lambda \zeta))}{(\boldsymbol{\omega} \cdot \boldsymbol{\nu}_\lambda^0 + \sigma_\lambda \zeta)^2}, \quad (5.4)$$

where $\sigma_\lambda = 1$ if $\lambda$ is on the *resonance path* $Q_V$, ($\lambda \in Q_V$), else $\sigma_\lambda = 0$.

REMARKS. (1) In writing (5.3) we take into account that one has $\chi^2(x) = \chi(x)$, for all $x \in \mathbb{R}$. Then we can associate to the external lines $\lambda_V$ and $\lambda'_V$ of a resonance $V$, the factors in square brackets in (5.3). Obviously the same argument can be repeated each time two lines have the same momentum and the same scale.
(2) The notion of resonance path makes sense, and therefore will be introduced, also for virtual resonances, although we shall consider it only in §5.7 below.

Let $\chi^{(n,n')}(x)$ denote the characteristic function of the set $|x| \in [2^{n-1}, 2^{n'})$; we see that the sum of all the scale values consistent with the maximal real resonances $\mathbf{V}_1$ yields as a result that (5.3) is identical with the expression in which the values $\mathcal{V}$ are modified as follows:

$$\mathcal{V}'(\zeta | V, \{n_\lambda\}_{\lambda \in V}) = \prod_{\substack{\lambda \in V \\ \lambda = xy}} \boldsymbol{\nu}_x \cdot \boldsymbol{\nu}_y \frac{\chi^{(n_{\lambda_V}+3, +\infty)}(\boldsymbol{\omega} \cdot \boldsymbol{\nu}_\lambda^0 + \sigma_\lambda \zeta)}{(\boldsymbol{\omega} \cdot \boldsymbol{\nu}_\lambda^0 + \sigma_\lambda \zeta)^2}, \quad (5.5)$$

having held fixed the scales of the lines outside the real resonances in $\mathbf{V}_1$.

We now develop (5.5) in powers of $\zeta$ and consider the second order remainder of the Taylor expansion around $\zeta = 0$. This quantity is given, by using Lagrange's interpolation, by:

$$\begin{aligned}
\zeta^2 \int_0^1 dt_V \, (1 - t_V) \frac{\partial^2}{\partial t_V^2} & \Big[ \prod_{\substack{\lambda \in Q_V \\ \lambda = xy}} \boldsymbol{\nu}_x \cdot \boldsymbol{\nu}_y \frac{\chi^{(n_{\lambda_V}+3, +\infty)}(\boldsymbol{\omega} \cdot \boldsymbol{\nu}_\lambda^0 + t_V \zeta)}{(\boldsymbol{\omega} \cdot \boldsymbol{\nu}_\lambda^0 + t_V \zeta)^2} \Big] \cdot \\
& \cdot \Big( \prod_{\substack{\lambda \in V/Q_V \\ \lambda = xy}} \boldsymbol{\nu}_x \cdot \boldsymbol{\nu}_y \frac{\chi^{(n_{\lambda_V}+3, \infty)}(\boldsymbol{\omega} \cdot \boldsymbol{\nu}_\lambda^0)}{(\boldsymbol{\omega} \cdot \boldsymbol{\nu}_\lambda^0)^2} \Big) .
\end{aligned} \quad (5.6)$$

We consider, temporarily, the value $X$ of the graph with the real resonance value corresponding to $V \in \mathbf{V}_1$ simply replaced by the expression defined in (5.6); this will be motivated in §5.7 below.

We can perform explicitly the derivatives in (5.6) and use that the characteristic functions yield, by differentiation, functions proportional to delta functions or their derivatives.

REMARK. Many terms cancel or just vanish. In particular various $\delta$-functions which end up being evaluated at zero cancel as one checks by suitably integrating by parts the terms containing two derivatives acting on the characteristic functions. Strictly speaking one needs also that $\boldsymbol{\omega} \cdot \boldsymbol{\nu} \neq 2^n$ for all $n \in \mathbb{Z}$, $\mathbf{0} \neq \boldsymbol{\nu} \in \mathbb{Z}^\ell$: but this condition is in fact not necessary at all if some ill defined expressions, like products of characteristic functions times delta functions, are discussed in detail: alternatively one can proceed as in [GG] in solving the similar problem that arose in [G2], where Kolmogorov's theorem was discussed under a somewhat stronger diophantine condition.

And after some time the following expression for the auxiliary quantity that we just called, following (5.6), $X$ materializes:

$$X = (-1)^k \Big( \prod_{v \in \hat{\vartheta}} f_{\nu_v} \Big) \cdot \Big( \prod_{\substack{\lambda \in \hat{\vartheta} \\ \lambda = xy}} \boldsymbol{\nu}_x \cdot \boldsymbol{\nu}_y \Big) \cdot \Big( \prod_{\lambda \in \hat{\vartheta}/\mathbf{V}_1} \frac{\chi(2^{-n_\lambda}(\boldsymbol{\omega} \cdot \boldsymbol{\nu}_\lambda))}{(\boldsymbol{\omega} \cdot \boldsymbol{\nu}_\lambda)^2} \Big) \cdot$$

$$\cdot \sum_{\lambda_1, \lambda_2 \in Q_V} \sum_{z=0}^{2} \int_0^1 dt_V \, (1-t_V) \, p(\lambda_1, \lambda_2, z, t_V) \cdot \qquad (5.7)$$

$$\cdot \Big\{ \prod_{V \in \mathbf{V}_1} \Big[ \prod_{\lambda \in V/Q_V} \frac{\chi^{(n_{\lambda_V}+3, +\infty)}(\boldsymbol{\omega} \cdot \boldsymbol{\nu}_\lambda)}{(\boldsymbol{\omega} \cdot \boldsymbol{\nu}_\lambda)^2} \cdot \prod_{\lambda \in Q_V} \frac{\chi^{(n_{\lambda_V}+3, +\infty)}(\boldsymbol{\omega} \cdot \boldsymbol{\nu}_\lambda(t_V))}{(\boldsymbol{\omega} \cdot \boldsymbol{\nu}_\lambda(t_V))^2} \Big] \Big\},$$

where we adopt the notations $\boldsymbol{\nu}_\lambda^0 + t_V \boldsymbol{\nu}_{\lambda_V} = \boldsymbol{\nu}_\lambda(t_V)$, and denote $p(\lambda_1, \lambda_2, z, t_V)$ the quantity $p$ defined by:

$$\begin{aligned} p &= (\boldsymbol{\omega} \cdot \boldsymbol{\nu}_{\lambda_V})^2 \begin{cases} 4 \frac{1}{\boldsymbol{\omega} \cdot \boldsymbol{\nu}_{\lambda_1}(t_V)} \frac{1}{\boldsymbol{\omega} \cdot \boldsymbol{\nu}_{\lambda_2}(t_V)}, & \text{if } z = 2, \lambda_1 \neq \lambda_2, \\ 6 \frac{1}{(\boldsymbol{\omega} \cdot \boldsymbol{\nu}_{\lambda_1}(t_V))^2}, & \text{if } z = 2, \lambda_1 = \lambda_2, \end{cases} \\ p &= (\boldsymbol{\omega} \cdot \boldsymbol{\nu}_{\lambda_V})(-2) \sum_{t_V^*} \delta(t_V - t_V^*) \frac{1}{\boldsymbol{\omega} \cdot \boldsymbol{\nu}_{\lambda_2}(t_V)}, & \text{if } z = 1, \\ p &= \delta_{\lambda_1, \lambda_2} \sum_{t_V^*} \delta(t_V - t_V^*) \frac{1}{1 - t_V}, & \text{if } z = 0, \end{aligned} \qquad (5.8)$$

where $t_V^*$ are the solutions to the equation $|\boldsymbol{\omega} \cdot \boldsymbol{\nu}_{\lambda_1}(t_V)| = 2^{n_{\lambda_V}+2}$, if any (there are at most 2 solutions). All the other (in principle) possible terms different from (5.8) cancel exactly by integrating by parts, as it can be easily checked, (being careful to take into account also the remark after (5.6)). The label $z$ denotes the number of derivatives acting on divisors.

We then suppose to redecompose, in (5.7), (5.8), the characteristic functions of the lines inside the real resonances into individual scales from $n_{\lambda_V} + 3$ up, by writing $\chi^{(n_{\lambda_V}+3, \infty)}(\cdot) = \sum_{n_{\lambda_V}+3}^{0} \chi(2^{-n} \cdot) + \chi_1(\cdot)$.

We recall that $\{n_\lambda\} \& \mathbf{V}_1$ denotes that the set of scales $\{n_\lambda\}$ are consistent with the existence of the family of real resonances $\mathbf{V}_1$ (see (5.3)). Then note that *if we could forget the first two orders of the Taylor expansions* that we have disregarded in (5.6) (by applying the renormalization procedure described above), then the sum of the values of the graphs of order $k$ with a given $\hat{\vartheta}$, i.e. the l.h.s. of (5.3), would be given by a formula that can be immediately read from (5.7), (5.8). Since we have forgotten the first two orders we get instead a quantity that we can call $Y_1$.

However to pursue the analysis it is very convenient to try to write the result $Y_1$ in a form as close as possible to (5.3), or to the original (3.3). After some meditation on the best way to express such complicated expressions as those in (5.7), (5.8), the sum of the quantity $X$ over the scales $\{n_\lambda\}$ consistent with $\mathbf{V}_1$, i.e. $Y_1$, is given by:

$$\sum_{\mathbf{V}_1} \sum_{\{n_\lambda\} \& \mathbf{V}_1} \Big( \prod_{v \in \hat{\vartheta}} f_{\nu_v} \Big) \Big[ \Big( \prod_{V \in \mathbf{V}_1} \int_0^1 dt_V \, (1-t_V) \Big) \cdot \prod_{\substack{\lambda \in \hat{\vartheta}, \lambda = xy \\ \lambda \notin \{\lambda_V'\}_{V \in \mathbf{V}_1}}} \frac{\chi(2^{-n_\lambda} \boldsymbol{\omega} \cdot \boldsymbol{\nu}_\lambda(\mathbf{t}))}{(\boldsymbol{\omega} \cdot \boldsymbol{\nu}_\lambda(\mathbf{t}))^2} \cdot$$

$$\cdot \Big( \prod_{\substack{\lambda \\ \lambda = xy}} \boldsymbol{\nu}_x \cdot \boldsymbol{\nu}_y \Big) \cdot \Big( \prod_{V \in \mathbf{V}_1} \sum_{\lambda_1^V, \lambda_2^V \in Q_V} \sum_{z^V=0}^{2} d_{\lambda_1^V \lambda_2^V}^{z^V} \frac{1}{\boldsymbol{\omega} \cdot \boldsymbol{\nu}_{\lambda_1^V}(\mathbf{t})} \cdot \frac{1}{\boldsymbol{\omega} \cdot \boldsymbol{\nu}_{\lambda_2^V}(\mathbf{t})} \Big) \Big], \qquad (5.9)$$

where $\mathbf{t} = \{t_V\}_{V \in \mathbf{V}_1}$ and we set $\boldsymbol{\nu}_\lambda(\mathbf{t}) = \boldsymbol{\nu}_\lambda^0 + t_V \boldsymbol{\nu}_{\lambda_V}$ if $\lambda \in Q_V$, and $\boldsymbol{\nu}_\lambda(\mathbf{t}) = \boldsymbol{\nu}_\lambda^0 \equiv \boldsymbol{\nu}_\lambda$ if $\lambda \notin \cup_{V \in \mathbf{V}_1} Q_V$, and:

$$\begin{aligned}
d^2_{\lambda_1 \lambda_2} &= 4\,\delta_{\lambda_1 \neq \lambda_2} + 6\,\delta_{\lambda_1 \lambda_2}\,, \\
d^1_{\lambda_1 \lambda_2} &= -2\,\frac{\boldsymbol{\omega} \cdot \boldsymbol{\nu}_{\lambda_1}(\mathbf{t})}{\boldsymbol{\omega} \cdot \boldsymbol{\nu}_{\lambda_V}} \sum_{t_V^*} \delta(t_V - t_V^*)\,, \\
d^0_{\lambda_1 \lambda_2} &= \delta_{\lambda_1, \lambda_2} \frac{(\boldsymbol{\omega} \cdot \boldsymbol{\nu}_{\lambda_1}(\mathbf{t}))^2}{(\boldsymbol{\omega} \cdot \boldsymbol{\nu}_{\lambda_V})^2} \sum_{t_V^*} \delta(t_V - t_V^*) \frac{1}{1 - t_V^*}\,,
\end{aligned} \qquad (5.10)$$

where $\delta_{\lambda_1 \neq \lambda_2} = 1 - \delta_{\lambda_1, \lambda_2}$. We call each addend in (5.9), with fixed $\vartheta = (\hat{\vartheta}, \{n_\lambda\})$ and $\{\lambda_1^V, \lambda_2^V, z^V, t_V\}_{V \in \mathbf{V}_1}$, the value of the graph $(\hat{\vartheta}, \{n_\lambda\})$ *superficially renormalized* on the real resonances $\mathbf{V}_1$, on the pairs of lines $\lambda_1^V, \lambda_2^V$ and on the choices $z^V$. There may be choices of $\{n_\lambda\}$ which are not compatible with the existence of the family of real resonances $\mathbf{V}_1$: in such cases the corresponding addends have (of course) to be interpreted as 0.

The expression in (5.9), once the summation labels $\vartheta$ and $\{\lambda_1^V, \lambda_2^V, z^V, t_V\}$ have been fixed, *i.e.* the value of the graph $\vartheta$ superficially renormalized on $\{\lambda_1^V, \lambda_2^V, z^V, t_V\}$ just defined, is clearly formally very close to the (3.3) we started with.

Note that the cases $z^V = 0, 1$ are special as they force $n_{\lambda_1^V} = n_{\lambda_V} + 3$ (which yields also $n_{\lambda_2^V} = n_{\lambda_V} + 3$, for $z^V = 0$): we say that in this case there is *no large scale jump* between the scale $n_{\lambda_V}$ entering the real resonance and the resonance scale $n_V$ (in fact the jump is exactly 3). The characteristic function $\chi(2^{-n_{\lambda_1^V}} \boldsymbol{\omega} \cdot \boldsymbol{\nu}_{\lambda_1^V}(\mathbf{t}))$, in the definition of the terms in (5.9) that involve it, is not necessary and it has been introduced to uniformize the notation (note that it has necessarily value 1 if the delta function is considered).

The absence of a large scale jump (or more properly a scale jump of 3 units) in the cases $z^V = 0, 1$ implies that the ratio of divisors in the definition (5.10) of $d^2_{\lambda_1 \lambda_2}$ is bounded above by $2^{n_{\lambda_V}+3}/2^{n_{\lambda_V}-1} = 2^4$. We see that although the cases $z^V = 0, 1$ do not give a factor $(\boldsymbol{\omega} \cdot \boldsymbol{\nu}_{\lambda_V})^2$ but only $\boldsymbol{\omega} \cdot \boldsymbol{\nu}_{\lambda_V}$, (see (5.8), $z = 1$), or just 1, (see (5.8), $z = 0$), the net result is essentially the same (up to the factor $2^{4(2-z^V)}$) because $\boldsymbol{\omega} \cdot \boldsymbol{\nu}_{\lambda_V}$ and $\boldsymbol{\omega} \cdot \boldsymbol{\nu}_{\lambda_1}(t_V)$ have a bounded ratio, since the latter is fixed by the delta function to be $2^{n_{\lambda_V}+2}$.

**5.4. Full renormalization.** Having dealt with the maximal real resonances (*first generation real resonances*) we perform again the same operations: *i.e.* fixed $\hat{\vartheta}$, $\mathbf{V}_1$, $\{\lambda_1^V, \lambda_2^V, z^V, t_V\}_{V \in \mathbf{V}_1}$ and the scales $\{n_\lambda\}$ for $\lambda \notin \cup_{V \in \mathbf{V}_1} V$, we identify the *second generation* real resonances as the maximal real resonances inside each $V \in \mathbf{V}_1$; call $\mathbf{V}_2$ the set of the real resonances of the first *and* second generations and proceed in a similar way to "renormalize" superficially the newly considered real resonances $W \in \mathbf{V}_2/\mathbf{V}_1$.

This means that we fix the scale labels of the lines outside the two generations of real resonances, and sum over the other scale labels $\{n_\lambda\}$ consistent with the elements of $\mathbf{V}_2$ being the first and second generation resonances.

We obtain that the product in (5.9) of the terms coming from the lines $\lambda \in W \in \mathbf{V}_2$, $W \subset V \in \mathbf{V}_1$ can be written in a form very close to (5.5), or (3.3), with the difference that the momenta flowing through the lines $\lambda \in Q_W \cap Q_V$ are: $\boldsymbol{\nu}_\lambda(\mathbf{t}) = \boldsymbol{\nu}_\lambda^0 + t_W(\boldsymbol{\nu}_{\lambda_W}^0 + t_V \boldsymbol{\nu}_{\lambda_V})$. And $n_{\lambda_V} + 3$ is replaced by $n_{\lambda_W} + 3$ in the characteristic functions. There are "just" a few extra labels, *i.e.* the summation labels over $V, \lambda_1^V, \lambda_2^V, z^V$ and a few factors due to the $d^{z^V}_{\lambda_1 \lambda_2}$ coefficients, $V \in \mathbf{V}_1$.

We proceed to do a Taylor expansion as above, in the variables $\zeta_W = \boldsymbol{\omega} \cdot (\boldsymbol{\nu}_{\lambda_W}^0 + t_V \boldsymbol{\nu}_{\lambda_V})$ if $\lambda_W \in Q_V$ or $\zeta_W = \boldsymbol{\omega} \cdot \boldsymbol{\nu}_{\lambda_W}^0$ otherwise. However this time we modify the procedure according to the number of lines $\lambda_1^V, \lambda_2^V$ that are in $W$: if $W$ contains both lines we do nothing. If $W$ contains one line we do the Taylor expansion "stopping at first order", *i.e.* we write the first order remainder; if $W$ contains none among $\lambda_1^V, \lambda_2^V$ we proceed as above and write the second order remainder.

We then perform the derivatives with respect to the new interpolation parameters $t_W$, generated by the expression of the Taylor series remainders and perform the cancellations of most of the terms involving the characteristic functions derivatives of first and second order (when the latter is present). Finally redevelop the characteristic functions and rearrange, along the lines that generated (5.9), the various terms to simplify the notation and to get an expression very similar to (5.9), see below, for a quantity that we could call $Y_2$.

The latter $Y_2$ can subsequently be used, in the same way as the $Y_1$ was already used to start the second renormalizations, for the superficial renormalization of the third generation of real resonances.

Then we iterate step by step the procedure until there are no more real resonances inside the maximal real resonances found at the last step performed and all the $n_\lambda$ have been fixed.

The final expression, that we can call $Y$, *can still be written* as (5.9) with $\mathbf{V}_1$ replaced by $\mathbf{V}$, which is the collection of all the real resonances selected along the procedure and with some of the integrals over the interpolation variables $t_V$ performed with weight $(1-t_V)$ and others with weight 1 (because some interpolations have been done to construct the first order remainder rather than the second order one), see §5.5 for details.

The (5.9) interpreted in this way (see below for a detailed description) will be called the *fully renormalized value* of the graph $(\hat{\vartheta}, \{n_\lambda\})$ and it can be denoted as $\mathcal{R}\mathrm{Val}(\vartheta)$.

**5.5. Details on full renormalization.** It is sufficient to state the result of the iteration described above. Given $\vartheta = (\hat{\vartheta}, \{n_\lambda\})$ let $\mathbf{V}$ be the set of real resonances and for $V \in \mathbf{V}$ let $\lambda_V', \lambda_V$ be the lines exiting and entering the real resonance.

We define, to abridge notations:

$$P_0(\vartheta) = \prod_{\lambda \notin \cup_V \lambda_V'} \frac{\chi(2^{-n_\lambda}\boldsymbol{\omega} \cdot \boldsymbol{\nu}_\lambda(\mathbf{t}))}{(\boldsymbol{\omega} \cdot \boldsymbol{\nu}_\lambda(\mathbf{t}))^2}, \qquad N(\vartheta) = \prod_{\substack{\lambda \in \hat{\vartheta} \\ \lambda = xy}} \boldsymbol{\nu}_x \cdot \boldsymbol{\nu}_y, \qquad (5.11)$$

where $\mathbf{t} = \{t_V\}_{V \in \mathbf{V}}$ are interpolation variables which eventually have to be integrated with respect to some measure $\pi_V(t_V) dt_V$, and $\boldsymbol{\nu}_{\lambda_v}(\mathbf{t})$ is a suitable (to be described later, see (5.13) below) linear combination of the external momenta $\boldsymbol{\nu}_w$ for $w \leq v$ with coefficients that are products of the interpolation variables $t_Z$ for the real resonances $Z$ that contain $\lambda$.

For each $V \in \mathbf{V}$ define a pair of lines $\lambda_1^V, \lambda_2^V \subset Q_V$, *i.e.* on the real resonance path of $V$, with the "compatibility condition" that, if $V$ is inside some other real resonances $Z$, then $\{\lambda_1^V, \lambda_2^V\}$ must contain the lines of the set $\cup_{Z \supset V}\{\lambda_1^Z, \lambda_2^Z\}$ that fall in $V$ which, therefore, can be 0, 1 or 2 at most. In the case the latter number is 1 we suppose that the lines have been labeled so that this is so because of the second line $\lambda_2^V$, (*i.e.* $\lambda_2^V \in \{\lambda_1^Z, \lambda_2^Z\}$). We call the three cases (0), (1), (2), respectively:
• if case (0) is realized for the pair $\lambda_1^V, \lambda_2^V$ we say that the lines $\lambda_1^V, \lambda_2^V$ are "new" and that the real resonance $V$ is new,
• in case (1) we say that the first line is new and the second "old" and that the real resonance is "partially new", and
• in the third case (2) both lines and the real resonance are old.
(The name is due to the order of appearance of the lines in the iteration steps).

Furthemore for each $V$ we define a variable $z^V = 0, 1, 2$ and, depending on which case among (0), (1), (2) above is realized, the functions $d^{z^V}_{\lambda^V_1 \lambda^V_2}$:

$$\begin{cases} d^2_{\lambda^V_1 \lambda^V_2} = 4\,\delta_{\lambda_1 \neq \lambda_2} + 6\,\delta_{\lambda^V_1, \lambda^V_2}\,, \\ d^1_{\lambda^V_1 \lambda^V_2} = -2 \frac{\boldsymbol{\omega} \cdot \boldsymbol{\nu}_{\lambda^V_1}(\mathbf{t})}{\boldsymbol{\omega} \cdot \boldsymbol{\nu}_{\lambda_V}(\mathbf{t})} \sum_{t^*_V} \delta(t_V - t^*_V)\,, \\ d^0_{\lambda^V_1 \lambda^V_2} = \frac{(\boldsymbol{\omega} \cdot \boldsymbol{\nu}_{\lambda^V_1}(\mathbf{t}))^2}{(\boldsymbol{\omega} \cdot \boldsymbol{\nu}_{\lambda_V}(\mathbf{t}))^2} \sum_{t^*_V} \delta(t_V - t^*_V) \frac{1}{1 - t^*_V}\,, \end{cases} \quad \text{if (0)}\,, \\ \begin{cases} d^2_{\lambda^V_1 \lambda^V_2} = -3\delta_{\lambda_1, \lambda_2} - 2\delta_{\lambda_1 \neq \lambda_2}\,, \qquad d^0_{\lambda^V_1 \lambda^V_2} = 0\,, \\ d^1_{\lambda^V_1 \lambda^V_2} = \frac{\boldsymbol{\omega} \cdot \boldsymbol{\nu}_{\lambda^V_1}(\mathbf{t})}{\boldsymbol{\omega} \cdot \boldsymbol{\nu}_{\lambda_V}(\mathbf{t})} \sum_{t^*_V} \delta(t_V - t^*_V)\,, \end{cases} \quad \text{if (1)}\,, \\ d^2_{\lambda^V_1 \lambda^V_2} = 1\,, \qquad d^1_{\lambda^V_1 \lambda^V_2} = 0\,, \qquad d^0_{\lambda^V_1 \lambda^V_2} = 0\,, \quad \text{if (2)}\,, \tag{5.12}$$

where $t^*_V$ are the solutions (at most 2) of the equation $|\boldsymbol{\omega} \cdot \boldsymbol{\nu}_{\lambda^V_1}(\mathbf{t})| = 2^{n_{\lambda^V}+2}$ for $t_V$.

We shall denote by $\Lambda$ the function $V \to \{\lambda^V_1, \lambda^V_2, z^V\}$ and the interpolation measures will be:
- $\pi_V(dt_V) = (1 - t_V)dt_V$ if $V$ is new,
- $\pi_V(dt_V) = dt_V$ if $V$ is partially old, and
- $\pi(t_V) = \delta(t_V - 1)dt_V$ if $V$ is old.

In (5.12) the interpolated momenta $\boldsymbol{\nu}_\lambda(\mathbf{t})$ are defined as follows. For each line $\lambda = w'w$ we consider all the real resonances $W$ that contain it on the respective resonance path $Q_W$ and define $t(w, \mathbf{t})$ to be the product of all the interpolation variables $t_W$ of such real resonances and $t(w, \mathbf{t}) = 1$ if there are no such resonances. The interpolated momenta are then defined by:

$$\boldsymbol{\nu}_{\lambda_v}(\mathbf{t}) = \sum_{w \leq v} t(w, \mathbf{t})\,\boldsymbol{\nu}_w\,. \tag{5.13}$$

Then we define:

$$\mathcal{R}\,\text{Val}(\vartheta) = \sum_\Lambda \prod_{V \in \mathbf{V}} \int_0^1 \pi(dt_V)\,\mathcal{R}\,\text{Val}(\vartheta, \{\lambda^V_1, \lambda^V_2, z^V, t_V\})\,, \tag{5.14}$$

where, see (5.11):

$$\mathcal{R}\,\text{Val}(\vartheta, \{\lambda^V_1, \lambda^V_2, z^V, t_V\}) = P_0(\vartheta)N(\vartheta) \prod_{V \in \mathbf{V}} d^{z^V}_{\lambda^V_1, \lambda^V_2} \left[ \frac{1}{(\boldsymbol{\omega} \cdot \boldsymbol{\nu}_{\lambda^V_1}(\mathbf{t}))^*} \frac{1}{(\boldsymbol{\omega} \cdot \boldsymbol{\nu}_{\lambda^V_2}(\mathbf{t}))^*} \right]\,, \tag{5.15}$$

and the $^*$ means that the two divisors have a different meaning from what has been written unless the real resonance $V$ is new. More precisely:
- if the real resonance is new, then $(\boldsymbol{\omega} \cdot \boldsymbol{\nu}_{\lambda^V_1}(\mathbf{t}))^* = \boldsymbol{\omega} \cdot \boldsymbol{\nu}_{\lambda^V_1}(\mathbf{t})$ and $(\boldsymbol{\omega} \cdot \boldsymbol{\nu}_{\lambda^V_2}(\mathbf{t}))^* = \boldsymbol{\omega} \cdot \boldsymbol{\nu}_{\lambda^V_2}(\mathbf{t})$,
- if the real resonance is partially old then $(\boldsymbol{\omega} \cdot \boldsymbol{\nu}_{\lambda^V_2}(\mathbf{t}))^*$ (corresponding to the "old" line $\lambda^V_2$) should be $\boldsymbol{\omega} \cdot \boldsymbol{\nu}_{\lambda_V}(\mathbf{t})$, and
- if both lines are old then both $(\boldsymbol{\omega} \cdot \boldsymbol{\nu}_{\lambda^V_1}(\mathbf{t}))^*$ and $(\boldsymbol{\omega} \cdot \boldsymbol{\nu}_{\lambda^V_2}(\mathbf{t}))^*$ should be $\boldsymbol{\omega} \cdot \boldsymbol{\nu}_{\lambda_V}(\mathbf{t})$.

One checks that the result of the above described iteration is that the sum over the scale labels $\{n_\lambda\}$ of the quantity in (5.14) coincides with the l.h.s. of (5.3), *with the new definition (5.5) of resonance value*.

The number of terms thus generated is, at fixed $\mathbf{V}$, bounded by the product over $V \in \mathbf{V}$ of 2 times the number of pairs that are in $V/ \cup_{W \subset V, W \in \mathbf{V}} W$ and therefore it is bounded by $2^k \prod_V k(V)^2$ if $k(V)$ is the number of nodes in $V$ which are not in real resonances inside $V$. Hence this number is $\leq (2^4)^k$.

The number of families of real resonances in $\hat{\vartheta}$ (*hence at fixed* $\{\boldsymbol{\nu}_x\}$) is also bounded by $2^k$.

We must deal, to find a complete expression of the sum of the values in the l.h.s. of (5.3), with the terms discarded in each Taylor expansion.

The key point is of course the following statement.

**LEMMA.** *Only graphs with real resonances whose factors are modified by the $\mathcal{R}$ operation have to be considered since the errors committed in so doing (due to having "forgotten" the various terms of order 0 or 1 in the Taylor expansions retaining only remainders), when all kth order graphs are summed together, including the summation over the momenta, give a vanishing contribution.*

See [GM1], or [GM3], §6, §7, for the proof. This is essentially the infrared cancellation remarked in [E]: its use here is slightly different from the use made in [G2].

Therefore (5.14) can be used as an exact expression for the calculation of the sum of the graph values.

The present procedure, introduced in [GM1], is conceptually different from the one in [G2], although the latter is perhaps more natural and, in the polynomial case, it seems to give better final estimates (this is due to the fact that the notion of real resonance in [G2] is more restrictive, so that the procedure we use here is "oversubtracting").

We say, interpreting the relevant features of the above discussion, that the application of $\mathcal{R}$ is realized by adding at most $2^{4k}$ terms each obtained by first eliminating two out of the four identical divisors corresponding to the lines entering and exiting the real resonances and then by "doubling" two divisors (possibly identical) among those in the part $V_0$ of each real resonance $V$ formed by the lines which are in $V$ but not in inner real resonances, as prescribed by the appropriate labels in (5.15), and finally "interpolating". There is, in some sense, an exception: there are some special cases (*i.e.* $V$ such that $z^V = 0, 1$ above) in which either only one or no factor $\boldsymbol{\omega} \cdot \boldsymbol{\nu}_{\lambda_V}(\mathbf{t})$ is really gained by the renormalization (*i.e.* one could say that "one of the two divisors of the external lines is left"), but in such cases there is *no big variation of scale* between the line incoming the real resonance and the scale of the lines inside the resonance (it is $= 3$).

One can also say that the $\mathcal{R}$ operation "eliminates" two of the four equal divisors correponding to the lines entering and exiting a real resonance and "transforms" the divisors of the two exiting lines into a pair of divisors for lines inside the real resonance. This will be sufficient once we try to get bounds.

The reader familiar with [GM3] will wonder why we have not taken the infrared decomposition in (5.1) as built with smooth functions as in [GM3]. This would have been indeed possible and in fact the argument would have been easier as no derivatives of characteristic functions would have ever arisen.

Although the amount of work would be somewhat reduced as one does not have to deal with delta functions (some of which end up being evaluated at 0 in the present approach and have to cancel) the use of characteristic functions is, in our opinion, a more pure approach and checking it in detail is a check of consistency. Furthermore the use of sharp characteristic functions in (5.1) is extremely useful in the theory that has been developed in [GGM] and in formulating the related conjectures; so we felt that it would be useful to have the details explicitly written (they are only sketched in [GM3]).

**5.6. Infrared bound.** The analysis of §5.4 allows us to find a bound on the sum over $n_\lambda = 1, 0, -1, \ldots$ of the values of the various graphs *after* the resonant factors have been modified by the action of the operations $\mathcal{R}$, leading to (5.14).

After applying the $\mathcal{R}$ operations, we see that the contribution to the new "renormalized value" from the divisors in (5.9) will be bounded by the same product appearing in the non renormalized values of the graphs *deprived of the divisors due to the lines exiting resonances* times a factor, see (5.14) and (5.11):

$$2^{4k} \prod_{V \subset \vartheta} \frac{1}{\min_{\lambda \in V_0}(\boldsymbol{\omega} \cdot \boldsymbol{\nu}_\lambda(\mathbf{t}))^2} \leq C_1^k \prod_{V \subset \vartheta} \frac{1}{\min_{\lambda \in V_0}(\boldsymbol{\omega} \cdot \boldsymbol{\nu}_\lambda^0)^2} \leq$$
$$\leq C_1^k \prod_{V \subset \vartheta} \Big[ \sum_{v \in V_0} |\boldsymbol{\nu}_v| \Big]^{2\tau}, \tag{5.16}$$

where we denote by $V_0$ the set of nodes inside the real resonance $V$ not contained in the real resonances internal to $V$, and $\boldsymbol{\nu}_\lambda^0$ is the momentum (called above "intrinsic momentum") flowing through the line $\lambda \in V_0$, when $\boldsymbol{\nu}_{\lambda_V}$ is set equal to zero; $C_1$ is a suitable positive constant. We have also used that the scale jump between the scale of the lines incoming a real resonance and the ones inside it is at least 3 as this permits us to replace $\boldsymbol{\omega} \cdot \boldsymbol{\nu}_\lambda(\mathbf{t})$ by $\boldsymbol{\omega} \cdot \boldsymbol{\nu}_\lambda^0$ in the above chain of inequalities.

In fact if $\lambda \in V$ but $\lambda$ is not inside any inner real resonance then $\boldsymbol{\nu}_\lambda(\mathbf{t})$ differs from $\boldsymbol{\nu}_\lambda^0$ by a vector corresponding to the line (interpolated) momentum of the line $\lambda_V$ entering $V$ which, because of the scale fixing characteristic functions, has a scale $n_{\lambda_V}$ at least 3 units lower than the scale $n$ of $\boldsymbol{\nu}_\lambda^0 + \boldsymbol{\nu}_{\lambda_V}(\mathbf{t})$ in the unfavourable case, (when $\lambda \in Q_V$). In this case $|\boldsymbol{\omega} \cdot \boldsymbol{\nu}_\lambda(\mathbf{t})| \geq |\boldsymbol{\omega} \cdot \boldsymbol{\nu}_\lambda^0|(1 - 2^{n_{\lambda_V}}/|\boldsymbol{\omega} \cdot \boldsymbol{\nu}_\lambda^0|)$, and $|\boldsymbol{\omega} \cdot \boldsymbol{\nu}_\lambda^0| \geq 2^{n_{\lambda_V}+2} - 2^{n_{\lambda_V}}$ hence $|\boldsymbol{\omega} \cdot \boldsymbol{\nu}_\lambda(\mathbf{t})| \geq \frac{2}{3}|\boldsymbol{\omega}_\lambda \cdot \boldsymbol{\nu}_\lambda^0|$.

Therefore, coming back to the full expression (5.14), (5.15) and to make use of the above (5.16), we consider a contribution $\mathcal{R}\,\text{Val}(\vartheta, \{\lambda_1^V, \lambda_2^V, z^V, t_V\})$ of a graph $\vartheta = (\hat{\vartheta}, \{n_\lambda\})$ to (5.14), with fixed scale labels $\{n_\lambda\}$ and fixed interpolation labels $\{\lambda_1^V, \lambda_2^V, z^V, t_V\}_{V \in \mathbf{V}}$. We can again, see §5.5, identify in it real resonances $V \in \mathbf{V}$ of different *generations*. The set $\mathbf{V}_j$ of real resonances of the $j$th generation, $j \geq 1$, just consists of the real resonances which are contained in $(j-1)$th generation real resonances (of lower scale) but not in any $(j+1)$th generation real resonances.

If $V$ is a real resonance in $\mathbf{V}_j$ with entering line $v_b^V v_1^V$ and outgoing line $v_0^V v_a^V$ with momentum $\boldsymbol{\nu}_{\lambda_V}$ we can construct a "$V$-contracted graph" by replacing the cluster $V$ together with the incoming and outgoing lines by the single line $v_0^V v_1^V$: *i.e.* by deleting the resonance $V$ and replacing it by a line. We can also construct the "$V$-cut graphs" by deleting everything but the lines of the resonance $V$ and its entering and outgoing lines and, furthermore, by deleting the outgoing line as well as the node $v_a^V$ and attributing to the node $v_1^V$ an external momentum equal to the momentum flowing into the entering line in the original graph $\vartheta$: thus we get $p_{v_a^V}$ disconnected graphs (recall that $p_v$ is in fact the number of lines merging into a node).

We repeat the above two operations until we are left only with graphs $\vartheta_i$, $i = 1, 2, \ldots$ without real resonances, all allowed in the sense of §3. *It is clear* that the product $\prod_{\lambda \in \vartheta}(\boldsymbol{\omega} \cdot \boldsymbol{\nu}_\lambda(\mathbf{t}))^{-2}$ is the same as the $\prod_i \prod_{\lambda \in \vartheta_i}(\boldsymbol{\omega} \cdot \boldsymbol{\nu}_\lambda(\mathbf{t}))^{-2}$.

Then we imagine to delete as well the lines of the various $\vartheta_j$ which were generated by the old entering lines (not all $\vartheta_i$ contain such lines, but some do) and we call $\vartheta_i^0$ the graphs so obtained. By doing so we change the momenta flowing into the lines of the graphs $\vartheta_i$ by an amount which is either $\mathbf{0}$ or the old momentum $\boldsymbol{\nu}_{\lambda_V}(\mathbf{t})$ entering a real resonance $V$. Since by definition of real resonance the latter has scale at least 3 units lower than the minimal scale of the lines in $\vartheta_i$, we have seen that $|\boldsymbol{\omega} \cdot \boldsymbol{\nu}_\lambda(\mathbf{t})| \geq \frac{2}{3}|\boldsymbol{\omega} \cdot \boldsymbol{\nu}_\lambda^0|$, if $\boldsymbol{\nu}_\lambda^0$ is the intrinsic momentum of $\lambda$ (*i.e.* the momentum flowing through $\lambda$ in $\vartheta_i^0$, a concept already introduced above).

Therefore we see that the product $\prod_{\lambda \in \hat{\vartheta}}(\boldsymbol{\omega} \cdot \boldsymbol{\nu}_\lambda(\mathbf{t}))^{-2}$ is bounded by $2^k \prod_i \prod_{\lambda \in \vartheta_i^0}(\boldsymbol{\omega} \cdot$ $\nu_\lambda^0)^{-2}$ times a factor equal to the product $\prod^*(\boldsymbol{\omega}\cdot\boldsymbol{\nu}_\lambda(\mathbf{t}))^{-2}$ of the divisors corresponding to the lines *exiting* the real resonances (of any order).

But the analysis of the properties of the $\mathcal{R}$ operation shows that its effect is just to replace the $\prod^*(\boldsymbol{\omega}\cdot\boldsymbol{\nu}_\lambda(\mathbf{t}))^{-2}$ by (5.16). Hence we see that we can bound each product of divisors $\prod_{\lambda\in\vartheta_i^0}(\boldsymbol{\omega}\cdot\boldsymbol{\nu}_\lambda^0)^{-2}$ intervening in the evaluation of the graphs values *after the $\mathcal{R}$ operations on the real resonances have been performed* by:

$$\prod_{\lambda\in\vartheta_i^0}(\boldsymbol{\omega}\cdot\boldsymbol{\nu}_\lambda^0)^{-2} \leq C^{2k}\frac{\prod_{v\in\vartheta_i^0}|\boldsymbol{\nu}_v|^{\eta\tau}}{(\sum_{v\in\vartheta_i^0}|\boldsymbol{\nu}_v|)^{2\tau}}, \tag{5.17}$$

with $\eta = 6$, by using lemma in §5.2.

This means that, if there were no factors $\boldsymbol{\nu}_{v'}\cdot\boldsymbol{\nu}_v$ associated with the lines $\lambda_v$ and contributing to the evaluation of the graph values, then we could bound each addend in (5.14) (*i.e.* a term in (5.15)) by:

$$C_1^k C^{2k}\prod_{v\in\vartheta}\chi(2^{-n_{\lambda_v}}\boldsymbol{\omega}\cdot\boldsymbol{\nu}_{\lambda_v}(\mathbf{t}))|f_{\boldsymbol{\nu}_v}||\boldsymbol{\nu}_v|^{\eta\tau}, \tag{5.18}$$

and if $b > \ell + \eta\tau$ we would have easily the absolute convergence of the "renormalized series".

However the factors $\boldsymbol{\nu}_{v'}\cdot\boldsymbol{\nu}_v$ *are precisely* those that cause the *ultraviolet divergences* and, before really looking for good bounds, we must use the ideas of §4 to overcome the problem.

**5.7. Ultraviolet cancellations.** Given a graph $\vartheta$ we shall naturally denote it as $\vartheta = (\vartheta^0, \{n_\lambda\}, \{\boldsymbol{\nu}_x\})$ where $\{n_\lambda\}$ is the set of scales of the momenta flowing in the branches $\lambda$ of $\vartheta^0$, due to the external momenta $\{\boldsymbol{\nu}_x\}$. The graph $\vartheta^0$ carries no scales nor momentum labels.

Proceeding as in §4.3, we could define the notion of nodes out of order, the sets $\mathcal{B}_{1v_1}$, see (4.11), (4.12), and the transformations $U \in \mathcal{U}(\mathcal{B}_{1v_1})$, see (4.12), and proceed to the path representation as in (4.4).

However it may happen that a pair of successive nodes $vw$, $v > w$ has $vw$ on the path of a *real or virtual* resonance $V$. Then the change of variables $U \in \mathcal{U}(\mathcal{B}_{1v}(\mathbf{P}))$ constructs a graph $(\vartheta^0, \{\boldsymbol{\nu}_\lambda\}, \prod_{w\in\mathcal{B}_{1v}(\mathbf{P})}U_{vw}^{\sigma w}\{\boldsymbol{\nu}_x\})$ in which the line incoming into the resonance carries some momentum $-\boldsymbol{\nu}$ while the outgoing line carries a momentum $\boldsymbol{\nu}$: hence in the new graph the cluster $V$ is no longer a resonance; or, viceversa, it can happen that a virtual resonance becomes real.

To avoid this "interference between ultraviolet and infrared cancellations" we *must* modify the ultraviolet interpolation procedure with respect to the one followed in §4.

Namely if $Q_V$ is a "resonance path" (defined in §5.3) and if we set $\mathcal{Q} = \cup_{V\in\tilde{\mathbf{V}}}Q_V$ with $\tilde{\mathbf{V}}$ the set of all the resonances of $\vartheta$ then we define $\mathcal{B}_v$ as the set of nodes $w$ preceding $v$ but such that *the line $vw$ in not on the resonance paths $\mathcal{Q}$*.

Consider the *renormalized value* of a graph $\vartheta = (\vartheta^0, \{n_\lambda\}, \{\boldsymbol{\nu}_x\})$, *i.e.* the value of the graph once the real resonance factors have been modified as described in §5.3÷5.5 to remove the infrared divergences.

By definition it is given by a term in (5.14), $\mathcal{R}\,\text{Val}(\vartheta, \{\lambda_1^V, \lambda_2^V, z^V, t_V\})$, with the scale labels $\{n_\lambda\}$ and the interpolation labels $\{\lambda_1^V, \lambda_2^V, z^V, t_V\}_{V\in\mathbf{V}}$ fixed. And define the collection $\mathcal{P}_1$ of paths on $\hat{\vartheta}$ as in §4 but not performing any interpolation if the node $w$ is on a resonance path.

Likewise one constructs the sets $\mathcal{P}_j$, always refraining from performing interpolations that would lead to paths $P$ with the highest line on resonance paths. One ends up with the family $\mathcal{P}$ of "partial" pavements and, for each $\mathbf{P} \in \mathcal{P}$, with the sets $\mathcal{B}_v(\mathbf{P}), \mathcal{B}_{1v}(\mathbf{P})$, as in §4. The pavements are partial because the union of the paths forming $\mathbf{P}$ can fail to cover the graph as it will not cover the union $\mathcal{Q}$ of the resonance paths.

The consequence is that for all $\mathbf{P} \in \mathcal{P}$ the change of variables $U \in \mathcal{U}(\mathcal{B}_{1v}(\mathbf{P}))$ change a graph $(\vartheta^0, \{n_\lambda\}, \{\boldsymbol{\nu}_x\})$ into a new graph $(\vartheta^0, \{n_\lambda\}, \prod_{w \in \mathcal{B}_{1v}(\mathbf{P})} U_{vw}^{\sigma_w}\{\boldsymbol{\nu}_x\})$ with the *same resonant clusters* (virtual or real).

The ultraviolet cancellations are now performed by interpolation as in §4. If one is careful in taking into account that the sum of the external momenta entering the nodes of a resonance vanishes one can restrict the sum in (4.5) to the $z$'s that are outside the resonances preceding $w$, thus getting less terms: a property which could be useful when attempting at getting bounds.

REMARK. The above definitions imply that, given $(\vartheta^0, \{\boldsymbol{\nu}_\lambda\}, \{\boldsymbol{\nu}_x\})$ and a transformation $U$, associated with a $\mathbf{P} \in \mathcal{P}$ and a set of signs $\{\sigma_w\}$ (with $v \in M_h(\mathbf{P})$ and $w \in \mathcal{B}_v(\mathbf{P})$), the product of the divisors:

$$\prod_{\lambda \in \vartheta} \frac{\chi(2^{-n_\lambda} \boldsymbol{\omega} \cdot \boldsymbol{\nu}_\lambda(\mathbf{t}))}{(\boldsymbol{\omega} \cdot \boldsymbol{\nu}_\lambda(\mathbf{t}))^2} \tag{5.19}$$

is the same for $(\vartheta^0, \{\boldsymbol{\nu}_\lambda\}, \{\boldsymbol{\nu}_x\})$ and $\vartheta = (\vartheta^0, \{n_\lambda\}, \prod_{w \in \mathcal{B}_{1v}(\mathbf{P})} U_{vw}^{\sigma_w}\{\boldsymbol{\nu}_x\})$ (simply because the $U$-operations generate graphs with the *same* or *opposite* momenta flowing through the branches) and the resonances are the same sets of lines. The interpolation does not affect this property by the definition of the $\boldsymbol{\nu}_\lambda(\mathbf{t})$. The new definition of $\mathcal{B}_v(\mathbf{P})$ just fixes the cases when this would be false (of course there will be soon or later a price to pay for this way out, see below).

We follow the notations of §4.4 but with the new meaning of the sets $\mathcal{P}$, $M_h(\mathbf{P})$, $M_e(\mathbf{P})$ and $\mathcal{B}_{1v}(\mathbf{P})$, $v \in M_h(\mathbf{P})$, consistent with the new definition of $\mathcal{B}_v$ and, hence, of $\mathcal{B}_v(\mathbf{P})$ (which cannot contain paths with highest line lying on a resonance path, since one "gives up" trying to use the parity cancellations that would correspond to such paths). By using also (5.16), we deduce a relation very close to (4.34), following the same considerations. With the notations in (4.34) this is, calling $\mathcal{R}W(\vartheta^0, \boldsymbol{\nu}) = \sum_{\{\boldsymbol{\nu}_x\}, \nu(\vartheta) = \nu} \mathcal{R} \operatorname{Val}(\vartheta)$, :

$$\sum_{\boldsymbol{\nu}} |\boldsymbol{\nu}|^s \Big| \sum_{\vartheta^0} \mathcal{R}W(\vartheta^0, \boldsymbol{\nu}) \Big| \leq$$

$$\leq C_0^k \sum_{\boldsymbol{\nu}} |\boldsymbol{\nu}|^s \Big| \sum_{\{n_\lambda\}} \sum_{\{h_x\}} \sum_{\mathbf{P} \in \mathcal{P}} \sum_{\{\boldsymbol{\nu}_\lambda\}}^* \Big[ \prod_{v \in \vartheta} |\boldsymbol{\nu}_v|^{\eta \tau} \Big] \cdot$$

$$\cdot \prod_{v \in M_h(\mathbf{P})} \Big( \prod_{w \in \mathcal{B}_{1v}(\mathbf{P})} \int_0^1 dt_w \Big) \sum_{||\mathbf{a}_v|| = p_v + 1} \mathcal{O}_v \Big( (i\boldsymbol{\nu}_v(\mathbf{t}_v))^{\mathbf{a}_v} Y_v(\mathbf{t}_v) f_{\boldsymbol{\nu}_v(\mathbf{t}_v)} \Big) \Big|, \tag{5.20}$$

with $\eta = 6$, by the same arguments described in §4, see (4.34) and by (5.18): here $\{n_\lambda\}$ is the set of infrared scales of $\{\boldsymbol{\nu}_\lambda\}$ and $\{h_x\}$ is the set of ultraviolet scales of the corresponding node momenta $\{\boldsymbol{\nu}_x\}$ (uniquely determined by the branch momenta $\{\boldsymbol{\nu}_\lambda\}$). As in §4 the sum $\sum_{\{\boldsymbol{\nu}_\lambda\}}^*$ is a sum over "representatives" but in the bounds we shall partially drop this constraint keeping only that $\{h_x\}$ is the set of ultraviolet scales of the momenta $\{\boldsymbol{\nu}_x\}$ (a property of the representatives discussed in §4). It will also be true, as in §4, that $2^{h_v - 1} \leq |\boldsymbol{\nu}_v| < 2^{h_v}$ and $2^{h_v - 2} \leq |\boldsymbol{\nu}_v(\mathbf{t}_v)| < 2^{h_v}$, (see (4.21)), so that we can obtain dimensional bound by the same mechanism already discussed in §4.

Note that not all the graphs involved by the ultraviolet cancellation have the same external momenta, (see (4.14)). Nevertheless they have all the same or opposite branch momenta (see the last remark), so that one can bound the small divisors product by $\prod_{v\in\vartheta}|\boldsymbol{\nu}_v|^{\eta\tau}$, *where the external momenta $\{\boldsymbol{\nu}_x\}$ are the ones of any graph among those between which there are the ultraviolet cancellations*: for instance one can and will choose the external momenta determined by the representatives.

The constant $C_0$ in (5.20) contains a product of the various constants that we built in §5.4, §5.5. Of course a trace remains of the cancellations: it is accounted by the factor $C_0^{\prime k}$ which, as mentioned above contains an estimate of the number $2^k$ of families of resonances $\mathbf{V}$, at fixed $\{\boldsymbol{\nu}_x\}$, see comments following (5.15).

The algebra necessary to bound (5.20) is the same as that of §4: there is however one obvious change: namely in the inequality corresponding to (4.40) we shall be left with an extra product $\prod_{vw\in\mathcal{Q}} 2^{h_v-h_w}$. This is so simply because we have *not* performed the subtractions relative to the lines with upper endodes belonging to $\mathcal{Q}$. Hence we deduce from the scaling properties of the $f_{\boldsymbol{\nu}_v}$, from lemma in §5.2, that (for a suitable $C_3$):

$$\sum_{\{\boldsymbol{\nu}\}}|\boldsymbol{\nu}|^s \Big|\sum_{\vartheta^0}\mathcal{R}W(\vartheta^0,\boldsymbol{\nu})\Big| \leq$$
$$\leq \max_{\mathbf{P}\in\mathcal{P}}\Big\{C_3^k\Big[\prod_{v\leq v_0} p_v!\Big]\sum_{\{h_x\}}\prod_{v\leq v_0}\Big[2^{h_v(1+p_v+s+\ell+\eta\tau-b)}\prod_{z:P(v,z)\in\mathbf{P}} 2^{(h_z-h_v)}\Big]\Big\}. \quad (5.21)$$

Then, setting $b = 2 + s + \ell + \eta\tau + \mu$, with $\mu > 0$, and exploiting an identity like (4.3), one obtains a bound on (5.20):

$$\sum_{\{\boldsymbol{\nu}\}}|\boldsymbol{\nu}|^s\Big|\sum_{\vartheta^0}\mathcal{R}W(\vartheta^0,\boldsymbol{\nu})\Big|\leq$$
$$\leq C^k\prod_v p_v!\sum_{\{h_x\}}\Big[2^{-(1+\mu)h_{v_0}}\prod_{v<v_0} 2^{-\mu h_v}\prod_{vw\in\mathcal{Q}} 2^{h_v-h_w}\Big], \quad (5.22)$$

for a suitable constant $C$. Thus everything is essentially identical to what was done in §4 except that this time the set $\mathcal{B}_v$ is smaller "than it should be" since we have not performed the cancellations on the nodes on resonance paths so that we "miss" the factor that would compensate the product $\prod_{vw\in\mathcal{Q}} 2^{h_v-h_w}$, which will not allow us to perform the summations on the scale labels (being very large if $h_v \gg h_w$.

However we see that *there is at most one factor $2^{h_v-h_w}$ per node $v$*, (because the resonance paths are totally ordered): hence we can easily compensate the unfavourable extra factors by requiring a slightly stronger condition on $b$, namely just one unit bigger: $b = 3 + s + \ell + \eta\tau + \mu$, with $\mu > 0$. With this assumption the $\mu$ is replaced by $1+\mu$ and the extra factors $2^{-h_v}$ compensate (when necessary) the factors $2^{h_v}$ in $2^{h_v-h_{v'}}$ (of course $2^{-h_{v'}} \leq 1$) that may lead to a divergence, so that the sum over the scale labels can be performed.

Again there are $k!/\prod_v p_v!$ graphs with given $p_v$'s and fixed shape so that the sum over the graph orders weighed by $\varepsilon^k$ can be performed if $\varepsilon$ is small enough; in particular we obtain that $\mathbf{h} \in C^{(s)}(\mathbb{T}^\ell)$, if $f \in \hat{C}^{(3+s+\eta\tau+\mu)}(\mathbb{T}^\ell)$, with $\mu > 0$. Thus the proof of Theorem 1.5 is complete.

28 *nov* 1995          31

## 6. Siegel-Eliasson's bound and further cancellations

The basic inequality (5.2) on products of the divisors in a graph, due to Eliasson extending a result by Siegel, is difficult to improve. We have made several attempts and we report here, without proofs, our results: we have not used them in this paper since in the end they do not improve Theorem 1.5 beyond what already obtained. Nevertheless they may have some interest in themselves, and they may stimulate work on clarifying which are the optimal bounds.

The bound in (5.2) is, however, likely to be not optimal, *e.g.* in [E] there is a statement that gives hopes that $6\tau$ can be reduced at least to $\frac{14}{3}\tau$.

A bound on the products of divisors which seems to indicate that $\eta\tau$ might be reduced is:

$$\prod_{\lambda \in \vartheta} \frac{1}{|\boldsymbol{\omega} \cdot \boldsymbol{\nu}_\lambda|} \leq k! B^k \prod_{v \in \vartheta} |\boldsymbol{\nu}_v|^\tau , \qquad (6.1)$$

for some $B$ and *for graphs which do not contain any connected subset $Z$ of nodes with* $\sum_{v \in Z} \boldsymbol{\nu}_v = \mathbf{0}$. We found for this inequality a simple proof.

Note that the above result (6.1) has some intrinsic interest: for instance in the case of the Siegel problem, [Pö], the $\boldsymbol{\nu}_v$'s have non negative components and therefore the conditions under which (6.1) holds are automatically fulfilled. The same can be said for the problem considered in [PV].

The inequality (6.1), at the beginning, gave us some hopes of allowing us to prove that the coefficients of the Lindstedt series are well defined uniformly in the ultraviolet cut off $\kappa > 0$ for $f \in \hat{C}^{(p)}(\mathbb{T}^\ell)$ for values of $p$ lower than the ones discussed above. See Theorem 1.5 where this result is claimed, *together with the analyticity at $\varepsilon$ small*, for $p > 6\tau + 3$. This would be an important first step towards an improvement of our results. But a closer analysis shows that the situation is quite delicate.

In fact starting from the proof of (6.1), obtained with methods close but not identical those of [S,E], we have introduced the notion of *weak resonance* as a cluster $W$ of connected nodes with:
(1) total momentum vanishing ($\sum_{v \in W} \boldsymbol{\nu}_v = \mathbf{0}$), and
(2) $s_W > 1$ incoming lines of scales which are $s_W + 2$ units lower than the cluster scale $n_W$.

Note that the condition (2) implies that there is necessarily one outgoing line. The notion of weak resonance that we give here is very different from the notions referred to as "resonances" with various qualifiers, in [E,CF], which always refer to situations in which two lines have equal momentum (which is not the case with the above weak resonances).

Then one can show, assuming (6.1) and following the ideas of the proof in Appendix A1 below, that a $k$-th order graph *without resonances and without weak resonances* has a product of divisors that can be bounded above by:

$$\prod_{\lambda \in \vartheta} \frac{1}{|\boldsymbol{\omega} \cdot \boldsymbol{\nu}_\lambda|^2} \leq b(k) \prod_{v \in \vartheta} |\boldsymbol{\nu}_v|^{4\tau} , \qquad (6.2)$$

for some $b(k)$ (that we estimate as bigger than $k!^2$).

Furthermore the notion of weak resonance allows us to use the existence of *other cancellations*, not used so far, among graphs containing clusters with $\mathbf{0}$ total momentum even when the clusters are not resonances in the sense of §5.2, but just weak resonances. Such cancellations (that are easy to see as they are due to the same mechanism discussed in [G2,GM1,GM3] in the case of the resonances) can be combined with an interpolation technique essentially identical to that of §5 and produce the same net effect found in §5, *i.e.* that the divisors of the lines outgoing from a weak resonance can be "transfered" on lines inside the resonance, as in the case of the resonances, see the interpretation of (5.15).

The factorials in the constants will certainly forbid getting a convergence proof: but the bound (6.2) combined with the methods of §5 might allow us to get finiteness to all orders of the Lindstedt series coefficients. Indeed it can be seen to give such result *but, unfortunately, still* under the condition that $p > 6\tau + 3$: in fact we get $4\tau$ from (6.2) and an extra $2\tau$ from the fact that two divisors inside a weak resonance are counted once more in the product (as in §5 with "ordinary resonances"), and a 3 for the same reasons as the 3 in §5. Hence "much ado about nothing": we work more and get a weaker result than Theorem 1.5.

Suppose that we insist, willing to forget convergence, in just trying to improve the value of $p$ that would at least guarantee that the sum rule for the Lindstedt series coefficients gives cofficients uniformly bounded, in the ultraviolet cut off $\kappa$, order by order.

It appears that we would need a better bound on the product of the squares of the divisors in (5.2) or, if we want to use the more sophisticated notion of weak resonance and the extra cancellations mentioned, a better bound of the squares of the divisors in a graph without weak resonances.

In fact if we could prove that the product of squares of divisors of a graph without weak resonances is bounded above, for suitable $k$–dependent constants $\Gamma_k$, by:

$$\Gamma_k \frac{\prod_{v \in \vartheta} |\boldsymbol{\nu}_v|^{4\tau}}{(\sum_{v \in \vartheta} |\boldsymbol{\nu}_v|)^{2\tau}} , \qquad (6.3)$$

(a bound that would considerably improve (6.2)), then, by the arguments in this paper, we could deduce that the Lindstedt series coefficients are at least finite, order by order, for $p > 4\tau + 3$.

But our attempts (whose details are not discussed here, as the results have not been used) at extending of the Siegel–Eliasson's bound only gave us (6.1), and (6.2) *without the denominator in (6.3)*: and this, we mentioned above, leads to no improvement on the results of this paper. A good enough justification for not reporting our proofs of (6.1) and of (6.2).

In any event (6.3), even if true, cannot be used for proving analyticity unless one will also be able to replace $\Gamma_k$ by $g^k$ for some $g$ (it "just" yields existence of the formal Lindstedt series). Hence, although we are not aware of counterexamples to (6.3), it is clear that the discussion cannot be based only on the hope that one can improve the lemma in §5.2 or use the cancellations that have not been taken into account.

## 7. Concluding remarks

**7.1. On regularity conditions.** A restriction, besides the parity property, with respect to the classical works [M1,M2,H] is the regularity requested on the function $f$, because the classical results only deal with the natural spaces $C^{(p)}(\mathbb{T}^\ell)$. Here the use of the spaces $\hat{C}^{(p)}(\mathbb{T}^\ell)$ seems essential for the above methods.

However it is clear that our method can be extended to rather more general functions.
For instance we could add, to any $f \in \hat{C}^{(p)}(\mathbb{T}^\ell)$ which verifies our assumptions, an arbitrary even analytic function. This means that we could replace the $|\boldsymbol{\nu}|^{-n}$ in (1.5) by any function of $\boldsymbol{\nu}$ which differs from it by an even quantity approaching 0 exponentially fast as $\boldsymbol{\nu} \to \infty$.

And we could replace $|\boldsymbol{\nu}|^{-n}$ in (1.5) by $P_i(\boldsymbol{\nu})|\boldsymbol{\nu}|^{-n-i}$ with $P_i(\boldsymbol{\nu})$ a homogeneous degree $i \geq 0$ even harmonic polynomial, see also [SW], p. 256, 282.

Another trivial generalization is to think the functions in (1.5) as sums of two functions of $|\boldsymbol{\nu}|$ which are homogeneous on two disjoint sublattices (the lattices of the even and odd $\boldsymbol{\nu}$'s, i.e. the lattices of the $\boldsymbol{\nu}$'s for which $(-1)^{|\boldsymbol{\nu}|} = +1$ or $(-1)^{|\boldsymbol{\nu}|} = -1$). The natural generalization is to consider a regular pavement of $\mathbb{Z}^\ell$ by cells, translates of a fundamental cell centered at the origin and reflection symmetric, and replace $|\boldsymbol{\nu}|^{-n}$ by $|\boldsymbol{\nu}|^{-n}$ times an arbitrary function $\pi(\boldsymbol{\nu})$ periodic in $\boldsymbol{\nu}$ with the period of the pavement and symmetric by reflection.

**7.2. On the parity condition.** The assumption that the Fourier transform of $f$ is even excludes interesting cases like $f_{\boldsymbol{\nu}} = i\nu_1|\boldsymbol{\nu}|^{-(b+1)}$. A different point of view seems necessary to include such cases in the theory: but the extension does not seem easy. By the remarks, and counterexamples, at the end of §2 we see that we cannot expect a simple extension to such cases and, in fact, we certainly do not expect, in general, analyticity in $\varepsilon$ near $\varepsilon = 0$ (see §2). However the problem is reminiscent of situations met already in quantum field theory (in the beta function theory, see [BG]) and it is not unlikely that something can be done also in this case.

**7.3. Best results.** Replacing, under appropriate conditions, the spaces $\hat{C}^{(p)}(\mathbb{T}^\ell)$ with wider subspaces of $C^{(p)}(\mathbb{T}^\ell)$ (in the even case) seems a harmonic analysis problem for which the techniques may be already known. An example of the conditions to which we think is that $\boldsymbol{\partial} f$ vanishes at the singularities of $f$ (to a large enough order). But it is likely that the method that we follow cannot lead to the "best results": the reason seems to be related to the exponent $\eta\tau$, with $\eta = 6$, in our extension of the Siegel-Eliasson's bound (lemma in §5.2) which is responsible for the $\eta\tau$ in the final result.

Restricting ourselves to the class of functions in $\hat{C}^{(p)}(\mathbb{T}^\ell)$ the discussion in §6 shows that our condition on $p$ (i.e. $p > 6\tau + 3$) is quite far from the best results on the Moser tori, see [H,M2]: in particular it is far from the classical result of Moser which yields the existence of invariant tori for $\varepsilon$ small under the condition $p > 2\tau + 4$, see [M2], last line, but not the analyticity in $\varepsilon$.

The results on analyticity might hold for $p \leq 6\tau + 3$, possibly just for $p > 2\tau + 4$: but the discussion in §6 shows that new ideas and techniques might be necessary.

Note, however, that such stronger statements may even fail to be true. As discussed above, in the present work we only get finiteness at all orders *and, simultaneously*, analyticity (i.e. convergence) for $p > 6\tau + 3$.

**7.4. A corollary.** The sum $\sum_{\boldsymbol{\nu} \neq \boldsymbol{0}} |\boldsymbol{\nu}|^{-\ell-s} e^{i\boldsymbol{\nu}\cdot\boldsymbol{\psi}}$ is a periodic function on $\mathbb{T}^\ell$ which is known to be $\gamma_s|\boldsymbol{\psi}|^{-s} + g_s(\boldsymbol{\psi})$ with $\gamma_s$ a suitable constant and $g_s$ a $C^{(\infty)}$ function for $|\psi_j| < \pi$, see [SW], p.282. Theorem 1.4 implies that $g$ is in fact real analytic for $\boldsymbol{\psi}$ such that $\partial_j f(\boldsymbol{\psi}) \neq 0$, $j = 1, \ldots, \ell$. A property that is not so easy to prove directly.

If one is willing to use the heavy proofs of this paper, this follows from the remark that our proof of Theorem 1.4 can be trivially adapted to the case in which the parameter $\varepsilon$ in the functional equation (1.4) is replaced by a diagonal matrix $(\varepsilon_1, \ldots, \varepsilon_\ell)$ of independent parameters. The solution is then analytic in $\varepsilon_j$ and $h_j$ is divisible by $\varepsilon_j$ (same proof of Theorem 1.4).

**7.5. Quantum field theory interpretation.** Finally we can comment that the field theoretic interpretation of the above constructions that we described in several papers ([G3,GGM]) on the analytic cases remains entirely valid: in fact the latter interpretation has been for us a guide to the proof of the above results, although one does not need it, and the field theory that it describes is not (yet) treatable by independent methods.

The field theory that corresponds (in the sense of [G3,GGM]) to the Moser's theorem is even more singular than the, already nasty, one corresponding to the analytic case: the action lagrangian is in fact the same function of the fields but this time it is only finitely many times differentiable. In the version in [G3] the field theory is described by two complex vector fields $\mathbf{F}^{\pm}_{\psi}$ on $\mathbb{T}^{\ell}$, with propagators given by:

$$\langle F^{+}_{\psi,j} F^{+}_{\psi',j'}\rangle = \langle F^{-}_{\psi,j} F^{-}_{\psi',j}\rangle = 0\ , \qquad \langle F^{+}_{\psi,j} F^{-}_{\psi',j'}\rangle = \delta_{ij} \sum_{\boldsymbol{\nu}\neq\mathbf{0}} \frac{e^{i\boldsymbol{\nu}\cdot\boldsymbol{\psi}}}{(i\boldsymbol{\omega}\cdot\boldsymbol{\nu})^2}\ , \qquad (7.1)$$

and the lagrangian is:

$$\mathcal{L}(\mathbf{F}) = \int_{\mathbb{T}^\ell} d\boldsymbol{\psi}\ \mathbf{F}^{-}_{\psi} \cdot \partial f(\boldsymbol{\psi} + \mathbf{F}^{+}_{\psi})\ . \qquad (7.2)$$

The connection with (1.3) is simply that the function $\mathbf{h}$ is given by the *one-point Schwinger function*:

$$\mathbf{h}(\boldsymbol{\psi}) = \frac{1}{Z} \int P(d\mathbf{F})\, e^{\varepsilon \mathcal{L}}\, \mathbf{F}^{+}_{\psi}\ , \qquad (7.3)$$

where $P(d\mathbf{F})$ denotes the (formal) gaussian integration with propagator ("covariance") (7.1) and $Z$ is the same functional integral with $\mathbf{F}^{+}$ replaced by 1 ("normalization").

A more general field theoretic interpretation which can be extended to cases in which $f$ is also $\mathbf{A}$–dependent, together with attempts at taking advantage of the field theoretic methods and ideas in order to understand properties of the *singularities* of $\mathbf{h}$ in $\varepsilon$ ("tori breakdown") can be found in [GGM] where it leads to some conjectures.

Hence the KAM theory has to be regarded as a technique to understand a field theory that has never arisen in the Physics literature and that, if treated via the usual techniques would look "untreatable" (*e.g.* it is *non renormalizable*): but a lot of progress in field theory has been achieved by discovering field theories that, although apparently untreatable (because non renormalizable (like gauge theories)) or even worse (*e.g.* conformal field theories), could be in fact treated and even sometimes "exactly solved".

**7.6. Hope.** Therefore one may hope for further applications on the field theory side.

## Appendix A1. Proof of Siegel-Eliasson's bound.

We prove here the lemma in §5.2: it only requires minor additions to the original argument by Eliasson. We call "product of divisors of a graph $\vartheta$" the product $\prod_{\lambda\in\vartheta} |\boldsymbol{\omega}\cdot\boldsymbol{\nu}_\lambda|^{-1}$.

Consider a graph $\vartheta$ without resonances in which there are two lines $\lambda_0 > \lambda_1$, $\lambda_0 = v_0 v_a$, $\lambda_1 = v_b v_1$, with the same momentum $\boldsymbol{\nu}$ of scale $n$. Two such lines will be called "closest" if there is not a line between $\lambda_0$ and $\lambda_1$ with scale $< n+3$ and if the subgraph $W$ formed by the lines joining the nodes $v \le v_a$ which are not below $v_1$
(i) does not contain other pairs of comparable lines with equal momentum, or
(ii) if it does contain such pair then the pair has an intermediate line $\tilde\lambda$ which has scale $< n' + 3$, if $n'$ is the scale of the pair of lines.

If there is no closest pair we deduce that for all comparable pairs there is an intermediate line with scale not higher than 2 units that of the pair.

Given a closest pair (if existent), the lines of $W$ different from $\lambda_1$ cannot form a cluster of scale $\ge n+3$ because we suppose that $\vartheta$ has no resonances.

Then $W$ contains a line $\tilde{\lambda}$, not comparable with $\lambda_1$, of scale $\leq n+2$, (all the lines on the path joining $\lambda_1$ to $\lambda_0$ have scale $\geq n+3$).

We cut out of $\vartheta$ two subgraphs: one is the already defined $W$ and the other is the graph $\tilde{\vartheta}$ obtained from $\vartheta$ by first deleting $W$ and then joining directly $v_0$ and $v_1$ by the line $\lambda_0$.

Break $W$ into subgraphs by deleting the node $v_a$: if $p_{v_a}$ is the number of branches merging into $v_a$ we obtain $p_{v_a}$ subgraphs that we denote $\tilde{\vartheta}_1, \vartheta_2, \ldots, \vartheta_{p_{v_a}}$. The graph $\tilde{\vartheta}_1$ is the one that still contains the line $\lambda_1$ and $v_1$ among the bottom nodes. We attribute to $v_1$, in $\tilde{\vartheta}_1$ as external momentum the momentum $\boldsymbol{\nu}$ of the line $\lambda_1$ in the original graph $\vartheta$ (note that in general $\boldsymbol{\nu}_{v_1} \neq \boldsymbol{\nu}$ in $\vartheta$).

It is clear that the product of the divisors of the graphs $\tilde{\vartheta}, \tilde{\vartheta}_1, \vartheta_2, \ldots$ is the *same* as that of the original graph $\vartheta$.

We now modify $\tilde{\vartheta}_1$ into a new graph $\vartheta_1$ by deleting the line $\lambda_1$ joining $v_b$ to $v_1$. In $\vartheta_1$ the divisors along the path that joins $v_b$ to the root of $\vartheta_1$ change with respect to what they were in $\tilde{\vartheta}_1$, from $\boldsymbol{\omega} \cdot (\boldsymbol{\nu}_\lambda^0 + \boldsymbol{\nu})$ to $\boldsymbol{\omega} \cdot \boldsymbol{\nu}_\lambda^0$. But since the lines of the path have scale 3 units higher than that of $\boldsymbol{\nu}$ we know that: $|\boldsymbol{\omega} \cdot (\boldsymbol{\nu}_\lambda^0 + \boldsymbol{\nu})| \geq \frac{2}{3}|\boldsymbol{\omega} \cdot \boldsymbol{\nu}_\lambda^0|$.

If we consider the graph among $\vartheta_1, \ldots, \vartheta_{p_{v_a}}$ which contains the line $\tilde{\lambda}$ with divisor $\delta'$ of scale $\leq n+1$, where $n$ is the scale of the divisor $\delta$ of the line $\lambda_1$, and mark it with a $*$ we see that the original product of divisors is bounded by the product of the divisors of $\tilde{\vartheta}, \vartheta_1, \ldots, \vartheta_j^*, \ldots, \vartheta_{p_{v_a}}$ provided in the graph bearing the $*$ label we square the divisor of the line with smallest scale, and multiply it by 8.

The graphs $\vartheta_1, \ldots, \vartheta_{p_{v_a}}$ are graphs without closest pairs of lines in the above sense. The graph $\tilde{\vartheta}$ may have such pairs and we can repeat on it the construction.

Thus we end up with a product of divisors associated with graphs that we call $\gamma_1, \gamma_2, \ldots$ that either have no pairs of comparable lines with equal momenta or may have such pairs, of scale $n$, with a line in between with scale $\leq n+2$. Some of the graphs $\gamma_i$ will be given a label $*$ to tell us that, at their mitosis from a larger graph, they "inherited" the divisor $\delta_i'$ of the line $\lambda_1$, not larger than $\frac{1}{8}$ the smallest diviisor of $\gamma_i$, that we call $\delta_i$. This means that in the bound on the product of divisors the original product is not larger than $\frac{1}{8}$ of the product of divisors of the graphs $\gamma_i$ in which the smallest divisor is squared.

The above discussion shows that we can restrict our attention to finding a bound for the graphs without pairs of lines with equal momentum or with such pairs separated along the graph by a line of scale at most 2 units higher.

A bound on the product of divisors of a graph without resonances is therefore closely related to a bound on the product of divisors of graphs of the latter type.

Eliasson has shown, see [E], §III, §IV, third lemma and related comments, that the product of the divisors in such a graph $\gamma$ of order $g$ can be bounded by:

$$AB^{g-1} \frac{\prod_{v \in \gamma} |\boldsymbol{\nu}_v|^{3\tau}}{(\sum_{v \in \gamma} |\boldsymbol{\nu}_v|)^{2\tau}}, \qquad (A1.1)$$

with $A, B$ suitably chosen constants.

Hence the product of divisors of the graphs $\gamma_i$ not marked by the $*$ are bounded (generously) by replacing $2\tau$ by $\tau$ in the denominator in (A1.1); the product of divisors for the starred graphs is bounded more carefully by using (A1.1) and by noting that the contribution of the extra divisor can be bounded above by $(\sum_{v \in \gamma} |\boldsymbol{\nu}_v|)^\tau$ compensated by the denominator in (A1.1) and still leaving a factor $(\sum_{v \in \gamma} |\boldsymbol{\nu}_v|)^\tau$ in the denominator. Hence the the product of divisors of the original graph is bounded by $A(8B)^{k-1} \frac{\prod_{v \in \vartheta} |\boldsymbol{\nu}_v|^{3\tau}}{(\sum_{v \in \vartheta} |\boldsymbol{\nu}_v|)^\tau}$, which gives (5.2).

## Appendix A2. On the number of pavements by ordered paths

We suppose first the case that a pavement covers the entire graph $\vartheta$. The case of a partial pavement (see §5.7) can be easily reduced to the first case.

We call $N(k)$ the maximum number of pavements of by disjoint ordered paths that can be drawn on a graph with $k$ branches and which have the highest node in one path. We shall check that $N(k) \leq 2^{k-1}$.

We shall assume first that the graph has no bifurcation at the first node (so that $k \geq 2$). For $k = 2$ it is $N(2) = 1$. We can suppose that $N(k) \leq 2^{k-2}$: then by induction we consider a graph with $k+1$ nodes and we see that either the path containing the highest node $v_0$ ends at the node $v_1$ immediately preceding it or it continues into one of the $p_0 \geq 1$ subgraphs $\vartheta_1, \ldots, \vartheta_{p_o}$ that have $v_1$ as root. Therefore, calling $k_1, \ldots, k_{p_0}$ the orders of the $p_0$ subgraphs, in the first case there are $\leq \prod_{j=1}^{p_0} N(k_j + 1)$ possible paths because we can think that the paths in the subgraphs $\vartheta_j$ are in fact paths on a graph which has one extra line preceding $v_1$ linking it to a new root auxiliary root $r_j$.

Likewise in the second case there are $p_0 \prod_{j=1}^{p_0} N(k_j + 1)$ possible paths (because there are $p_0$ possibilities to choose into which of the $p_0$ subgraphs the path containing $v_0 v_1$ will enter: and once we know that it enters $\vartheta_j$ then it has to be continued into a path among the $N(k_j)$ that are possible). Since $\sum_j k_j = k - 2$ we see that $N(k+1) \leq (1+p_0)2^{k-2-p_0} \leq 2^{k-1} \max_{x \geq 1}\{(1+x)2^{-1-x}\} = 2^{k-2}$.

In general a graph with $k$ nodes does not have a highest node $v_0$ which has no bifurcations: however we can add to a graph with $k$ nodes an extra node $\tilde{v}_0$, higher than $v_0$ and turn it into a $k+1$ nodes graph with a highest node without bifurcations. The number of paths on such a graph, with one path which starts at $\tilde{v}_0$, is $\leq 2^{k-1}$ of course (by the above inductive argument).

Note that the number of paths that can be drawn on a $k$-th order linear graph is exactly $2^{k-2}$: hence the above estimate is optimal for the graphs with a highest node without bifurcations.

One easily sees that in fact the above inequality (for the special graphs with a bifurcationless node before the root) implies that for a general graph $N(k) \leq 2^{-p_0} 2^{-(k-1)}$, if $p_0$ is the bifurcation of the node before the root. This is a result that can also be equally easily (in fact by an essentially identical argument) checked directly by induction.

It is easy to see that there are graphs for which $N(k)$ is exactly $2^{-p_0} 2^{-(k-1)}$ so that the estimate is optimal.

**Acknowledgements.** We are indebted to L.H. Eliasson and J. Moser for very helpful and enlightening discussions. This work is part of the research program of the European Network on: "Stability and Universality in Classical Mechanics", # ER-BCHRXCT940460.